\documentclass{jfm}
\usepackage{graphicx}
\usepackage{xcolor}
\usepackage{epstopdf,epsfig}
\usepackage{newtxtext}
\usepackage{newtxmath}
\usepackage{mathtools}
\usepackage{natbib}
\usepackage{hyperref}
\hypersetup{
    colorlinks = true,
    urlcolor   = blue,
    citecolor  = black,
}

\newcommand{\RomanNumeralCaps}[1]
\DeclareMathOperator{\erf}{erf}
\DeclareMathOperator{\erfc}{erfc}

\shorttitle{Linear stability analysis of a time-divergent slamming flow}
\shortauthor{Devaraj van der Meer}

\title{Linear stability analysis of a time-divergent slamming flow}
\author{Devaraj van der Meer
  \corresp{\email{d.vandermeer@utwente.nl}}}

\affiliation{Physics of Fluids Group and Max Planck Center Twente for Complex Fluid Dynamics, MESA+ Institute and J. M. Burgers Centre for Fluid Dynamics, University of Twente, P.O. Box 217, 7500AE Enschede, The Netherlands}

\begin{document}
\maketitle

\begin{abstract}
When a liquid slams into a solid, the intermediate gas is squeezed out at a speed that diverges when approaching the moment of impact. Although there is mounting experimental evidence that instabilities form on the liquid interface during such an event, understanding of the nature of these instabilities is limited. This study therefore addresses the stability of a liquid-gas interface with surface tension, subject to a diverging flow in the gas phase, where the liquid and the gas phase are both represented as potential fluids. We perform a Kelvin-Helmholtz-type linear modal stability analysis of the surface to obtain an amplitude equation that is subsequently analysed in detail and applied to two cases of interest for impact problems, namely, the parallel impact of a wave onto a vertical wall, and the impact of a horizontal plate onto a liquid surface. In both cases we find that long wavelengths are stabilised considerably in comparison to what may be expected based upon classical knowledge of the stability of interfaces subject to a constant gas flow. In the former case, this leads to the prediction of a marginally stable wavelength that is completely absent in the classical analysis. 
For the latter we find much resemblance to the classical case, with the connotation that the instability is suppressed for smaller disk sizes. The study ends with a discussion of the influence of gas viscosity and gas compressibility on the respective stability diagrams.        
\end{abstract}

%\begin{keywords}
%Authors should not enter keywords on the manuscript, as these must be chosen by the author during the online submission process and will then be added during the typesetting process (see \href{https://www.cambridge.org/core/journals/journal-of-fluid-mechanics/information/list-of-keywords}{Keyword PDF} for the full list).  Other classifications will be added at the same time.
%\end{keywords}
%
%{\bf MSC Codes }  {\it(Optional)} Please enter your MSC Codes here

\section{Introduction}
\label{sec:intro}

The importance of stability analysis for the development of our understanding of fluid flows can hardly be overstated. Instabilities of liquid-gas interfaces, of which the Kelvin-Helmholtz type may be one of the prime examples, have long found their way into textbooks \citep{Chandrasekhar1970,Drazin2002,Criminale2019}. They are of relevance to a great variety of different areas of physics, such as surface waves \citep{miles1959,nayfeh1972}, stratified shear flows \citep{peltier2003,funada2001}, mixing layers and turbulence \citep{rogers1992,smyth2000,zhou2017}, vortex sheets \citep{hou1997}, flow around objects \citep{magnaudet1995}, spray atomization \citep{beale1999}, meteorology, environmental and atmospheric physics \citep{fritts2003,ghisalberti2006}, magnetohydrodynamics \citep{dangelo1965,kent1969}, superfluids \citep{blaauwgeers2002,tsubota2013}, and astrophysics \citep{anderson2008,baiotti2017}.  

Nevertheless, the study of instabilities in time-dependent flows have received considerable less attention, and most  
research performed in that area has concentrated on oscillatory flows \citep{kelly1965,grosch1968,davis1976,poulin2003,troy2005,talib2007,yoshikawa2011}. A notable exception is the study of instabilities on falling viscous jets, which are advected by a non-uniform base flow and as such render the stability problem inherently time-dependent \citep{javadi2013}. In fact, we set out to study a very different type of time-dependent flow, namely one that is diverging. The reason for doing so stems from the context of impact problems. When a liquid slams into a solid, the gas phase that resides in the rapidly decreasing space between needs to be expelled at ever increasing speeds. Examples are the impact of breaking waves on structures \citep{peregrine2003,dias2018}, sloshing waves \citep{Faltinsen2009}, or the impact of a plate on a liquid surface \citep{abrate2011,kapsenberg2011,truscott2014,mayer2018,jain2020}. Especially in the context of overseas transport of cryogenic fuels, such as with liquid natural gas (LNG) carriers \citep{bogaert2019}, where the largest loads on the structure are know to be due to sloshing wave impact, it has recently become clear that instabilities that arise on breaking sloshing waves account for a significant part of the variability observed in sloshing impact loads \citep{lafeber2012a,lafeber2012b,bogaert2019}. Therefore, to accurately predict these impact loads it is crucial to understand the origin of these instabilities. This is precisely where the current work aims to contribute.    

All these impact events have in common that gas needs to be squeezed from the space between the impactor and the target. Since the liquid and the solid phases generally approach each other with an, at least in leading order, constant velocity $U_0$, the distance between impactor and target may be written as $d = U_0\tau$, where $\tau$ is equal to the amount of time remaining until impact. Now, continuity dictates that the ratio of the typical gas velocity $U_g$ and $U_0$ should be equal to the typical area of the impactor (or target, whichever is smallest) perpendicular to the direction of motion, divided by the product of the perimeter and the distance $d$. Since without loss of generality this area (and corresponding perimeter) may be taken to be constant, this implies that the gas velocity is inversely proportional to $\tau$: $U_g \sim 1/\tau$, which will serve as the basis of this work.   

As stated above, the main reason for this study comes from the impact of waves on structures, where a detailed understanding of the pressures that are generated during the impact is crucial to predicting the load that is experienced by the structure. Existing theories have evolved from those that mainly involve the impact of inertial liquids onto solids \citep{wagner1932,peregrine2003,korobkin2007} and which use potential flow theory as their main tool, to more recent ones that take into account the presence of the interstitial gas, either as a potential or a viscous fluid \citep{wilson1991,korobkin2008,hicks2013,bouwhuis2015,josserand2016}. However, all of these works deal with smooth impacting bodies of liquid, whereas the observation from the engineering practice is that --especially on the larger scale-- instabilities occur that change the shape of the impacting wave crest and consequently also change the load they exert on the impacted structure \citep{lafeber2012a,lafeber2012b,bogaert2019}. The main objective of this work is therefore to make a first step in predicting the wavelength and size of these instabilities, where due to the diverging nature of the gas flow we need to deal with a time-dependent version of classical linear stability analysis.

Now what would such a time-dependent stability analysis look like? First consider a classical Kelvin-Helmholtz instability of a flat, horizontal gas-liquid interface subject to a steady horizontal gas flow $U_g$. It is well-known (see, e.g., \cite{Drazin2002}) that if $U_g$ is larger than a certain limit (of marginal stability), the interface is unstable and in time a disturbance will appear with a certain wavelength corresponding to the mode with the largest growth rate. Now the first step to the time-dependent situation (which we will call the semi-classical case) is to assume that the gas velocity is very slowly increasing in time, such that the classical analysis can still be assumed to be valid. What would one observe in this case? At first, when $U_g$ is small, the interface is stable, and it remains stable until $U_g$ reaches the marginal stability threshold. Increasing $U_g$ a little bit beyond the threshold will cause the growth of an instability with the marginal wavelength of classical stability analysis, since that is the first one to become unstable. When $U_g$ increases even further, the instability has already occurred and even if the most unstable wavelength would change as a function of $U_g$ (it doesn't in this particular problem) this would have no large effect, since the interface is already deformed. The main task of a time-dependent stability analysis would therefore be to determine the wavelength that will become unstable first (which we will call the marginal wavelength $\lambda_\textrm{marg}$). The obvious path to accomplish this would be to determine the instability onset time ($\tau_\textrm{ons}$) for each of the modes and search for the earliest time, which we will call the marginal onset time ($\tau_\textrm{marg}$).

Now in the problem we aim to study, $U_g$ is definitely not always changing slowly in time, as it diverges close to impact. This implies that, next to the time scale that is dictated by the stability problem and determines how fast instabilities grow, there is the second time scale on which the gas flow $U_g$ is changing, and it is a priori unclear which of those two will be dominant. Therefore, we must formulate our stability analysis as a fully time-dependent problem and need to depart from the notion of exponential growth from classical linear stability analysis. We will find that the solutions are all growing as power-laws in the amount of time remaining until impact (i.e., in principle faster than exponential), but some of the modes will be calculated to grow slowly or even extremely slowly. To distinguish (extremely) slowly growing modes from others it is insufficient to determine when a mode becomes unstable, since this will give an unphysical answer for slow modes. We therefore need to distinguish modes while they are in the process of growing, which we will do by setting a threshold value $p$ and determining when the ratio of the amplitude and the initial amplitude of the disturbance grows beyond that value. This introduces an additional parameter to the analysis, but it is unavoidable to obtain physically meaningful results.

The paper is structured as follows. In Section~\ref{sec:problem} we will discuss the typical geometrical setup of the problem, introduce the basic equations of the modal stability analysis, and arrive at an amplitude equation. Subsequently, we will discuss properties and solutions of this equation in Section~\ref{sec:amplitude} from a more mathematical point of view and turn to their stability in connection with the concepts of the growth rate, the magnification factor and the threshold value $p$. We will also connect to classical linear Kelvin-Helmholtz stability theory and discuss in what respect the current work is different. In Section~\ref{sec:stability} we will then turn to the resulting stability diagrams for two physical cases of interest, namely the parallel impact of a wave onto a vertical wall and the impact of a horizontal plate onto a liquid surface, which constitutes the main physical result of this study. Subsequently, in Section~\ref{sec:viscous} we discuss the limiting influence of gas viscosity and compressibility on the stability analysis. Finally, we provide some tentative comparison to experimental results (Section~\ref{sec:comparison}), before concluding in Section~\ref{sec:conclusion}. To maintain structural clarity, some more technical points have been deferred to a series of Appendices.

\section{Problem statement and stability analysis}
\label{sec:problem}

\begin{figure}
 \centerline{\includegraphics[width=0.7\textwidth]{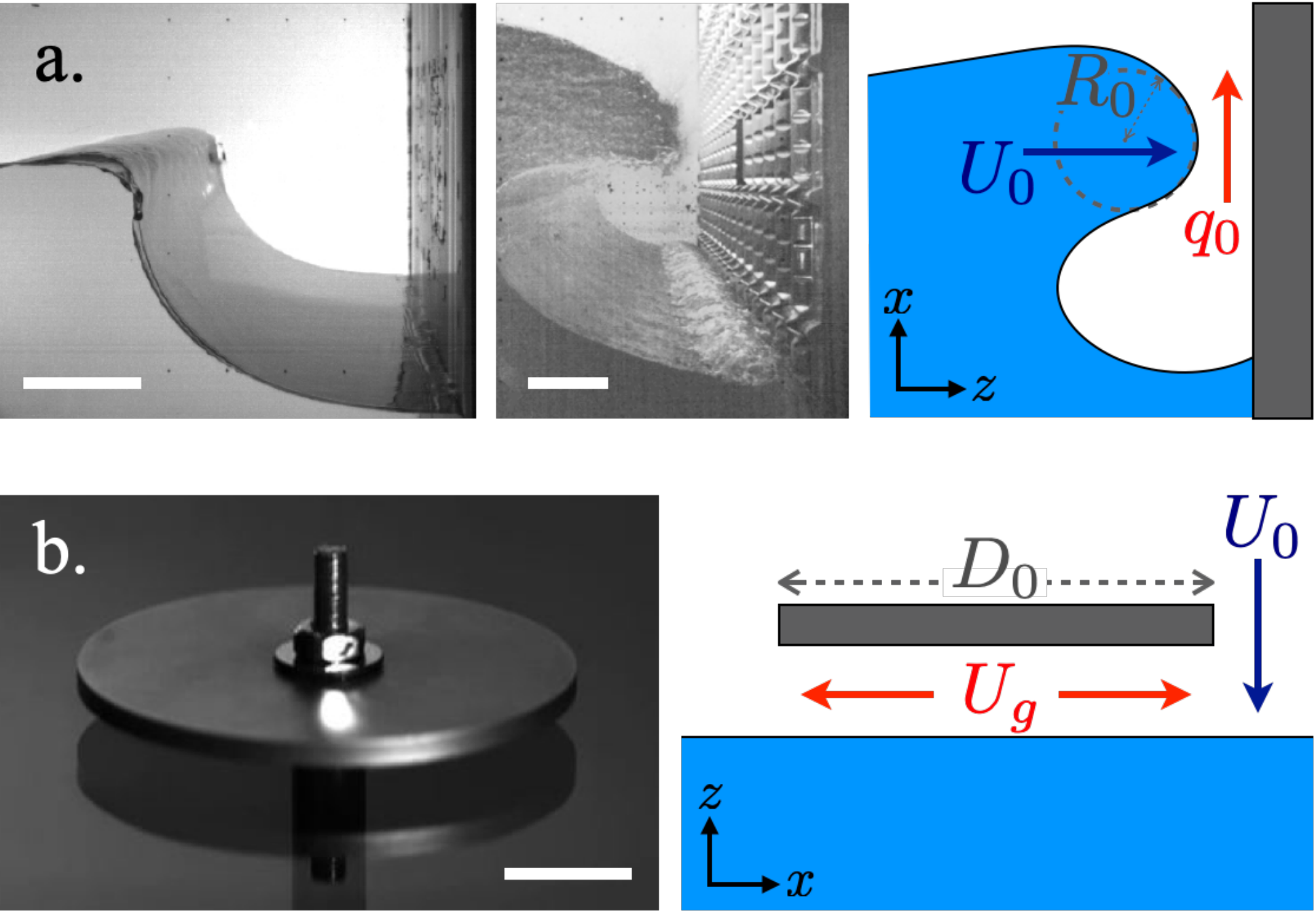}}% Images in 100% size
 \caption{Two types of slamming impact events studied in this work: a. The impact of a breaking wave onto vertical wall. Left: two pictures from wall impact experiments of breaking waves generated in two different flumes (taken from \cite{bogaert2019} with permission; the horizontal scale bars measure $10$ cm (left) and 20 cm (middle) respectively); right: a schematic of the gas flow during impact: the gas pocket volume changes at a constant rate, leading to a constant (two-dimensional) volumetric flow rate $q_0$ in the gap between crest and wall. Here, $U_0$ and $R_0$ are the velocity and local radius of curvature of the impacting crest, respectively.  b. The horizontal impact of a plate onto a water surface. Left: a circular disk is pulled through a water surface (the horizontal scale bar has a length of $1.0$ cm); right: a schematic of a circular disk of diameter $D_0$ impacting a water surface at a constant velocity $U_0$, which sets up a gas flow $U_g$ below the disk, which is maximal at the edge. In the two schematics the horizontal ($x$) and vertical ($z$) directions used in the analysis have been indicated for clarity.} 
\label{fig:slammingimpact}
\end{figure}

In this work we will concentrate on two types of slamming liquid-solid impact. The first is a quasi two-dimensional breaking wave, as is, e.g., created during sloshing, that slams into a vertical wall (Fig.~\ref{fig:slammingimpact}a). During such an impact, typically, a gas pocket gets entrapped below the wave. While the wave is moving towards the wall, the gas pocket decreases in size and consequently a gas flow is set up along the impacting crest. As the crest is approaching the wall, the width $d$ of the intermediate gap decreases to zero in time $t$. If the amount of time that remains until the moment of impact at time $t = t_i$, is denoted by $\tau = t_i - t$, the gap width equals $d= U_0\tau$, and the gas velocity $U_g$ in the gap can be written as
\begin{equation}\label{eq:gasflow}
U_g(\tau) = \frac{q_0}{U_0\tau}\,,
\end{equation}
where $q_0$ is the (two-dimensional) volumetric flow rate in the gap, which is equal to minus the rate of change of the gas pocket volume below the crest. It may be assumed to be constant, since the gap width is likely to change much more rapidly in time than the derivative of the gas pocket volume. Normally, the volume changes in the gas pocket do not lead to any compression or pressure increase of the gas inside the pocket. This will only happen if the escaping gas experiences large friction in the gap, which may be the case when either viscous or compressibility effects start to dominate the gas flow inside the gap. These effects will be studied in Section~\ref{sec:viscous}.  

The second type of slamming event is the impact of a horizontal circular disk of diameter $D_0$ impacting onto a liquid surface with a constant velocity $U_0$, as depicted in Fig.~\ref{fig:slammingimpact}b. Here, the gas needs to be pressed from the gap between the disk and the liquid surface, which using continuity and assuming a uniform radial gas flow in the gap leads to a gas velocity that depends on the radial coordinate, $U_g = \tfrac{1}{2}r/\tau$, and is maximal under the disk edge, where
\begin{equation}\label{eq:gasflow2}
U_g(\tau) = \frac{D_0}{4\tau} = \frac{q_0}{U_0\tau}\,.
\end{equation}
In the second step, we have cast the equation into the form of Eq.~\eqref{eq:gasflow} by defining the two-dimensional volumetric flow rate as $q_0 = U_0D_0/4$. Note that the same relation holds if the impacting object is a rectangular plate of width $W_0$, where the gas velocity under the plate edge is given by Eq.~\eqref{eq:gasflow} with $q_0 = U_0W_0/2$.
 
\begin{figure}
  \centerline{\includegraphics[width=0.95\textwidth]{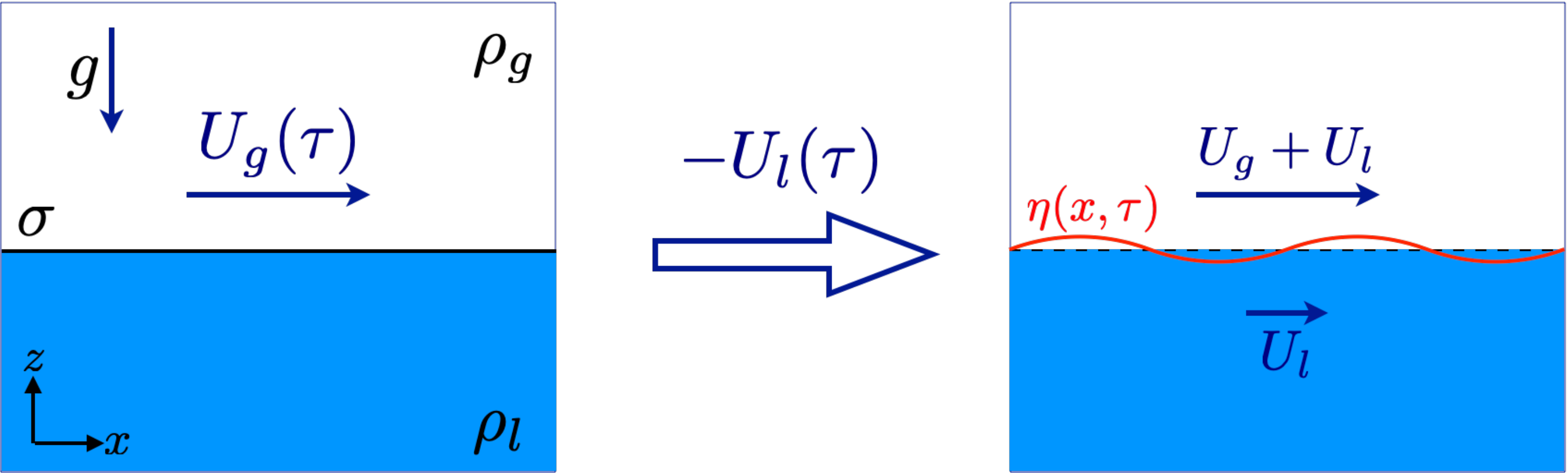}}% Images in 100% size
  \caption{Left: model setup for the stability analysis of an unsteady, diverging gas flow of uniform velocity $U_g(\tau)$ above a quiescent liquid, where the liquid-gas interface with surface tension $\sigma$ is positioned at $z=0$. The gas and liquid densities are denoted by $\rho_g$ and $\rho_l$ respectively, and the system is subject to a downward acceleration $g$ (e.g., due to gravity). Right: to create a base state that obeys the potential flow equations, the whole system needs to be viewed in an accelerated reference frame that moves at a velocity $U_l(\tau) = (\rho_g/(\rho_l-\rho_g))U_g(\tau)$ in the negative $x$-direction (see main text). In this reference frame, we will study the linear stability of the liquid-gas interface under a small perturbation $\eta(x,t)$. }
\label{fig:model}
\end{figure}

The above sketched slamming liquid impact geometries may be simplified to the following two-dimensional model setup, that is depicted in Fig.~\ref{fig:model} and captures the essence of the stability problem: We will assume a liquid of density $\rho_l$ and a gas of density $\rho_g$ with a liquid-gas interface (having interfacial tension $\sigma$), that is initially flat and above which there exists a uniform, time-dependent gas flow with velocity $U_g(\tau)$ of the form described in Eq.~\eqref{eq:gasflow} in the $x$-direction, parallel to the interface that lies in the horizontal plane. %$(x,y)$-plane. 
This will be the basic state to which we will perform a stability analysis within the context of potential flow by adding a disturbance $z = \eta(x,t)$. Although the analysis may be straightforwardly extended to a full, three-dimensional formulation (the result of which is provided in Appendix~\ref{appA} for completeness) we will restrict ourselves to small disturbances in the $x$-direction only, which is the direction in which the most unstable wave vectors will be found. Initially, our arguments closely follow the line of reasoning of the seminal paper of Kelly, who performed a similar stability analysis for oscillatory flows half a century ago \citep{kelly1965}.  

We start by noting that the situation depicted in the left schematic in Fig.~\ref{fig:model} cannot represent a valid basic flow for our analysis. The reason is that the pressure needs to be continuous at the interface, whereas if we use the bases state formed by $\phi_g^{(0)}  = U_g(\tau)x + F(t)$ and $\phi_l^{(0)} = G(t)$ for arbitrary functions of time $F$ and $G$, then the pressure condition at the interface leads to
\begin{equation}\label{eq:pressurecond_basic}
p_g|_{z=0} = p_l|_{z=0} \quad\Rightarrow\quad \tfrac{1}{2}\rho_gU_g^2 + \rho_g\frac{\partial U_g}{\partial t}x + \dot{F}(t) = \dot{G}(t)\,, 
\end{equation}
where we have used that the pressure jump due to surface tension is zero since the interface is flat and $\dot{F}$ and $\dot{G}$ denote the time derivatives of the functions $F$ and $G$. Clearly this condition can only be satisfied for time-independent $U_g$, 
%cannot be satisfied for any but constant $U_g$, 
because of the explicit $x$-dependence in the left hand side of Eq.~\eqref{eq:pressurecond_basic}. If however we translate the system with a time-dependent velocity $U_l(\tau)$ towards the negative $x$-direction, we obtain the system depicted in the right schematic of Fig.~\ref{fig:model}, where the liquid and gas obtain velocities $U_l$ and $U_g + U_l$ respectively, and the pressure condition \eqref{eq:pressurecond_basic} turns into
\begin{equation}\label{eq:pressurecond_basic_translated}
\tfrac{1}{2}\rho_g(U_g+U_l)^2 + \rho_g\frac{\partial (U_g+U_l)}{\partial t}x + \dot{F}(t) = \tfrac{1}{2}\rho_l(U_l)^2 + \rho_l\frac{\partial U_l}{\partial t}x + \dot{G}(t)\,.
\end{equation}
This equation may be satisfied for suitable $\dot{F}-\dot{G}$, provided that the $x$-dependent term vanishes, i.e., when
\begin{equation}\label{eq:liqspeedcond}
U_l(\tau) = \frac{\rho_g}{\rho_l - \rho_g} U_g(\tau) = \frac{\delta}{1 - \delta} U_g(\tau) \,,
\end{equation}
where we have introduced the gas-to-liquid density ratio $\delta = \rho_g/\rho_l$. Although the translated reference frame is non-inertial, it introduces %a fictitious 
an inertial acceleration in the $x$-direction only, which is not expected to interfere with the stability analysis that we are about to perform. Note that it does imply that the result of our stability analysis may be expected to translate with a velocity $U_l\approx\delta U_g$ towards the positive $x$-direction, i.e., with a velocity that is usually much smaller than the gas velocity $U_g$, as $\delta \ll 1$. (This will be addressed in more detail in Subsection~\ref{subsec:advectiveterms}.) 

It is good to stress the necessity of having a finite velocity in both phases in order for the basic flow to satisfy the potential flow equations, which necessarily leads to flow potentials of the form discussed above, and which for completeness are stated below
\begin{eqnarray}
\phi_l^{(0)}(x,t) = \frac{\delta}{1-\delta}U_g(\tau)x + F(t)\,,\nonumber\\
\phi_l^{(0)}(x,t) = \frac{1}{1-\delta}U_g(\tau)x + G(t)\,,\label{eq:baseflow}
\end{eqnarray}
where $F(t)$ and $G(t)$ are chosen to satisfy condition~\eqref{eq:pressurecond_basic_translated}

We now consider this basic flow to be perturbed by a small disturbance %of the basic flow 
that coincides with a small vertical displacement $z = \eta(x,t)$ of the flat interface at $z=0$. That is, there are perturbed flow potentials  $\phi_l  =  \phi_l^{(0)} +  \phi_l^{(1)}$ and  $\phi_g  =  \phi_g^{(0)} +  \phi_g^{(1)}$ in the liquid and the gas phase, which need to obey the continuity equation, the far field kinematic boundary conditions and the linearized kinematic and dynamic boundary conditions at the interface:
%\begin{eqnarray}
%\nabla^2 \phi_l &=& \nabla^2 \phi_l^{(1)} = 0\,,\\
%\nabla^2 \phi_g &=& \nabla^2 \phi_g^{(1)} = 0\,,\\
%\lim_{z\to-\infty}\phi_l^{(1)} &=& 0\,\\
%\lim_{z\to+\infty}\phi_g^{(1)} &=& 0\,\\
%\frac{\partial \eta}{\partial t} + U_l\frac{\partial \eta}{\partial x} &=& \frac{\partial \phi_l^{(1)}}{\partial z}\,,\\
%\frac{\partial \eta}{\partial t} + U_g\frac{\partial \eta}{\partial x} &=& \frac{\partial \phi_g^{(1)}}{\partial z}\,,\\ 
%\rho_l\left(\frac{\partial \phi_l^{(1)}}{\partial t\,\,}  + U_l \frac{\partial \phi_l^{(1)}}{\partial x\,\,} + g\eta \right) &=& \rho_g\left(\frac{\partial \phi_g^{(1)}}{\partial t\,\,}  + U_g \frac{\partial \phi_g^{(1)}}{\partial x\,\,}+ g\eta \right) + \sigma\frac{\partial^2 \eta}{\partial x^2}\,,
%\end{eqnarray} 
\begin{eqnarray}
\nabla^2 \phi_i = \nabla^2 \phi_i^{(1)} &=& 0\qquad\qquad(i=l,g)\,,\label{eq:Laplace}\\
\lim_{z\to\mp\infty}\phi_i^{(1)} &=& 0\qquad\qquad(i=l,g)\,,\label{eq:bottopBC}\\
\frac{\partial \eta}{\partial t} + \bar{U}_i\frac{\partial \eta}{\partial x} &=& \frac{\partial \phi_i^{(1)}}{\partial z}\qquad\,(i=l,g)\,,\label{eq:kinBC}\\
\rho_l\left(\frac{\partial \phi_l^{(1)}}{\partial t\,\,}  + \bar{U}_l \frac{\partial \phi_l^{(1)}}{\partial x\,\,} + g\eta \right) &=& \rho_g\left(\frac{\partial \phi_g^{(1)}}{\partial t\,\,}  + \bar{U}_g \frac{\partial \phi_g^{(1)}}{\partial x\,\,}+ g\eta \right) + \sigma\frac{\partial^2 \eta}{\partial x^2}\,,\label{eq:dynBC}
\end{eqnarray}   
where we have used that (up to quadratic order in $\eta$) all boundary conditions at the interface have been satisfied by the basic flow $ \phi_i^{(0)}$ and all derivatives of the $\phi_i^{(1)}$ may be evaluated at the undisturbed interface at $z=0$. Here, $i = l,g$ stands for either the liquid or the gas phase, $g$ is the acceleration component perpendicular to the interface  (e.g., gravity) and $\nabla^2$ represents the Laplacian. Note that for notational convenience we have defined $\bar{U}_l = U_l = (\delta/(1-\delta))U_g$ and $\bar{U}_g = U_g + U_l = (1/(1-\delta))U_g$ corresponding to the velocities in the two phases in the translated reference frame.

With a modal decomposition Ansatz for the displacement of the free surface, $\eta(x,t)=\varepsilon(t)\exp[ikx]$, we may solve the Laplace equations \eqref{eq:Laplace}, together with the far-field boundary conditions \eqref{eq:bottopBC} as $\phi_l^{(1)} = C_l(t)\exp[kz+ikx]$ and $\phi_g^{(1)} = C_g(t)\exp[-kz+ikx]$. Inserting these expressions in the kinematic \eqref{eq:kinBC} and dynamic \eqref{eq:dynBC} boundary conditions we obtain
\begin{eqnarray}
(\dot{\varepsilon} + ik\bar{U}_l\varepsilon) &=& kC_l\,,\label{eq:kbc1}\\
(\dot{\varepsilon} + ik\bar{U}_g\varepsilon) &=& -kC_g\,,\label{eq:kbc2}\\
\rho_l(\dot{C}_l + ik\bar{U}_lC_l +g\varepsilon) &=& \rho_g(\dot{C}_g + ik\bar{U}_gC_g +g\varepsilon) - \sigma k^2 \varepsilon\,. \label{eq:dbc}
\end{eqnarray}   
Multiplying the last equation \eqref{eq:dbc} with $k$ and substituting $kC_l$ and $kC_g$ from Eqs.~\eqref{eq:kbc1} and \eqref{eq:kbc2} leads to the following differential equation for the amplitude $\varepsilon(t)$
\begin{eqnarray}\label{eq:governing1}
(\rho_l+\rho_g)\ddot{\varepsilon} &+& 2ik(\rho_l\bar{U}_l+\rho_g\bar{U}_g)\dot{\varepsilon}\\
&+& \left[ik(\rho_l\dot{\bar{U}}_l+\rho_g\dot{\bar{U}}_g) - k^2(\rho_l\bar{U}_l^2+\rho_g\bar{U}_g^2) + (\rho_l-\rho_g)gk + \sigma k^3\right]\varepsilon =0\nonumber\,,
\end{eqnarray}
Now, following \cite{kelly1965}, we can get rid of the imaginary terms in the above equation by defining a modified amplitude $\bar{\varepsilon}(t)$ as
\begin{eqnarray}\label{eq:modified}
\bar{\varepsilon}(t) &=& \varepsilon(t)\exp\left[ik\int_{-\infty}^t\left(\frac{\rho_l}{\rho_l+\rho_g}\bar{U}_l(\tau') + \frac{\rho_g}{\rho_l+\rho_g}\bar{U}_g(\tau') \right)dt'\right]\nonumber\\
%=\varepsilon(t)\exp\left[ ik\frac{2\rho_l\rho_g}{\rho_l^2-\rho_g^2}\int_{-\infty}^t U_g(\tau')dt' \right]\,,
&=& \varepsilon(t)\exp\left[ ik\frac{2\delta}{1-\delta^2}\int_{-\infty}^t U_g(\tau')dt' \right] \equiv \varepsilon(t)\exp[ikH(t)]\,,
\end{eqnarray}
where we inserted $\bar{U}_l$ and $\bar{U}_g$ from their definitions to obtain the expression on the second line. Note that $|\bar{\varepsilon}(t)| = |\varepsilon(t)|$, implying that the above transformation does not alter the stability problem we seek to study. After inserting $\varepsilon = \bar{\varepsilon}\exp[-ikH]$ into Eq.~\eqref{eq:governing1} and some algebraic manipulation this leads to a modified amplitude equation for $\bar{\varepsilon}(t)$    
\begin{equation}\label{eq:governing2}
\frac{d^2\bar{\varepsilon}}{dt^2} + \left[\frac{1-\delta}{1+\delta}gk + \frac{\sigma}{\rho_l(1+\delta)}k^3 - \frac{\delta}{(1+\delta)^2}(U_g(\tau))^2 k^2\right]\bar{\varepsilon} = 0\,,
\end{equation}
again using the definitions of $\bar{U}_l$ and $\bar{U}_g$. Note that this result is almost identical to the one one would obtain in a traditional Kelvin-Helmholtz analysis~\citep{Drazin2002}, i.e., for constant $U_g$, with the exception of %the prefactor of the gas-velocity-dependent last term in the equation and the
fact that now the gas velocity depends on time. For the case that the gas velocity is given by Eq.~\eqref{eq:gasflow}, i.e., $U_g(\tau) = q_0/(U_0\tau)$, the latter will have large implications, as we will see in the next sections. 

We want to conclude the section with a short caveat, since the above simplification disregards many features that will inevitably be present in real systems. One obvious simplification is the assumption of a flat liquid interface, which for wave impact can only (very) approximately be realized, if at all. Another one is the assumed spatially uniform flow, which is impossible to exactly realize for both disk and rectangular plate impact, as the gas velocity will necessarily increase with the distance to the center as one approaches the rim, to not even speak of the spatially divergent character of the gas flow in the case of a circular disk. A third simplification is the description of terms of two infinite fluid half spaces, whereas in practice the gas-solid and gas-liquid interfaces approach each other, which at some point in time will violate the assumption of an infinitely large gas layer and may influence the stability of the free surface. In principle, these features could be included in a stability analysis along the same path that is followed here, albeit at the expense of the simple, analytic approach that is possible in this highly idealized description only, and that will be further elaborated upon in the next sections.

There are however a few more subtle issues, like the observation that during disk impact the liquid interface will be pushed slightly downward by the intervening gas layer, an effect known as air cushioning \citep{verhagen1967,wilson1991,peters2013,jain2020}. Thus, although the disturbance is typically very small, the interface is not entirely flat and the resulting acceleration could in principle modify the gravitational acceleration. Although one can show that the largest deformation takes place below the disk center and not at the rim where the instability should occur and that the acceleration typically remains small at time scales that are relevant for the current stability analysis, it is good to be alert to subtleties that may be of relevance in certain limits.

\section{Amplitude equation analysis}
\label{sec:amplitude}

We will start by discussing and characterizing the solutions of the amplitude equation~\eqref{eq:governing2}, which has the mathematical structure
\begin{equation}\label{eq:ampleq}
\frac{d^2y}{d\tau^2} + \left(A - \frac{B}{\tau^2}\right) y = 0\,,
\end{equation}
where we renamed $y=\bar{\varepsilon}$ and have used that $d\tau=-dt$, where we take the opportunity to remind the reader that $\tau = t_i-t$ is the amount of time remaining until impact, i.e., that we need to integrate Eq.~\eqref{eq:ampleq} backwards in $\tau$. We have introduced the positive constants $A$ and $B$ given by
\begin{eqnarray}
A(k) &\equiv& \frac{1-\delta}{1+\delta}gk + \frac{\sigma}{\rho_l(1+\delta)}k^3 \equiv \alpha_1 k + \alpha_3 k^3\,,\label{eq:Adef}\\
B(k) &\equiv& \frac{\delta}{(1+\delta)^2}\frac{q_0^2}{U_0^2} k^2 \equiv \beta_2k^2\,. \label{eq:Bdef}
\end{eqnarray} 
%%ALREADY DEFINE $w(\delta)$, THE PREFACTOR OF B(k) HERE!!!!!
Note that $A$ has the dimensions of inverse squared time whereas $B$ is dimensionless. Although, e.g., the key equation~\eqref{eq:ampleq} is conveniently expressed in terms of inverse time $\tau$, when discussing early and late events we will always refer to ordinary time $t$, i.e., events at a smaller $\tau$ occur later than events with larger $\tau$.   

%%MAYBE REFORMULATE THE NEXT SENTENCE IN THE LIGHT OF THE INTRODUCTION OF THE SEMI-CLASSICAL PICTURE LATER ON!!!
If we were to disregard the time-dependence of the third term in the above equation, we would be in the situation of classical Kelvin-Helmholtz stability theory, where a stabilizing term ($Ay$) competes with a destabilizing one ($-By/\tau^2$) and solutions would be purely oscillatory and stable when $A>B/\tau^2$ and unstable (in fact, exponentially diverging) when $B/\tau^2>A$. However, the fact that the third term is time dependent implies that we will typically see both types of behavior in a solution of Eq.~\eqref{eq:ampleq}: Initially, for large $\tau$, the solution will be (close to) a harmonic oscillator until, after reaching a threshold at $\tau \approx \sqrt{B/A}$, it will start to diverge.   

\subsection{Solution of the amplitude equation}
\label{subsec:amplitudesol}

Before looking in more detail at solutions, let us first observe that Eq.~\eqref{eq:ampleq} can be non-dimensionalised by introducing dimensionless time $\tilde{\tau} = \sqrt{A}\tau$ and amplitude $\tilde{y} = y/|y_0|$, where $y_0$ is the amplitude at some initial time $\tau_0 = t_i - t_0$, which then leads to the one parameter equation
\begin{equation}\label{eq:nondimampleq}
\frac{d^2\tilde{y}}{d\tilde{\tau}^2} + \left(1 - \frac{B}{\tilde{\tau}^2}\right) \tilde{y} = 0\,.
\end{equation}
Clearly, this equation has two distinct limits, namely an early time limit, $\tilde{\tau} \gg \sqrt{B}$, where the time-dependent last term is negligible and the equation reduces to that of a harmonic oscillator with dimensionless angular frequency 1 (corresponding to a dimensional angular frequency $\sqrt{A}$), and a late time limit, $\tilde{\tau} \ll \sqrt{B}$, just before impact at $\tilde{\tau} = 0$ where the last term in Eq.~\eqref{eq:nondimampleq} is dominant
\begin{eqnarray}
\frac{d^2\tilde{y}}{d\tilde{\tau}^2} + \tilde{y} &\approx& 0\qquad\textrm{for}\,\,\tilde{\tau} \gg \sqrt{B}\,,\label{eq:earlylimeq}\\
\frac{d^2\tilde{y}}{d\tilde{\tau}^2} - \frac{B}{\tilde{\tau}^2} \tilde{y} &\approx& 0\qquad\textrm{for}\,\,\tilde{\tau} \ll \sqrt{B}\,. \label{eq:latelimeq}
\end{eqnarray} 
Allowing for complex solutions, we may write the general solution to Eq.~\eqref{eq:earlylimeq} as 
\begin{equation}\label{eq:earlylimsol}
\tilde{y} \approx \exp[i\psi + i\tilde{\tau}] \qquad(\textrm{for}\,\,\tilde{\tau} \gg \sqrt{B})\,,
\end{equation}
where the prefactor of the exponential has been set to $1$, consistent with the definition of the dimensionless amplitude $\tilde{y}$, and $\psi$ is some constant initial phase. Also the second equation~\eqref{eq:latelimeq} can be solved straightforwardly by noting that one should try a power-law solution of the form $\tilde{y} \sim \tilde{\tau}^\kappa$, which when inserted into~\eqref{eq:latelimeq} leads to the algebraic relation $\kappa(\kappa-1) - B = 0$, which has solutions $\kappa_\pm =  \tfrac{1}{2}(1\pm\sqrt{4B+1})$ with which
\begin{equation}\label{eq:latelimsol}
\tilde{y} \approx \tilde{y}_+\tilde{\tau}^{\kappa_+} + \tilde{y}_-\tilde{\tau}^{\kappa_-} \approx \tilde{y}_-\tilde{\tau}^{\kappa_-}\qquad\textrm{with}\,\,\kappa_\pm =\tfrac{1}{2}(1\pm \sqrt{4B+1})\qquad(\textrm{for}\,\,\tilde{\tau} \ll \sqrt{B})\,.
\end{equation}
Clearly, as $\tilde{\tau}\to 0$, the solution is dominated by the second term, for which the exponent $\kappa_-$ is always negative. It is good to note that one may in fact also write down analytical solutions to Eq.~\eqref{eq:nondimampleq} in terms of Bessel functions. For completeness, these are provided in Appendix~\ref{appC}, but they are of little use for the purposes of this article.  

\begin{figure}
 \centerline{\includegraphics[width=0.85\textwidth]{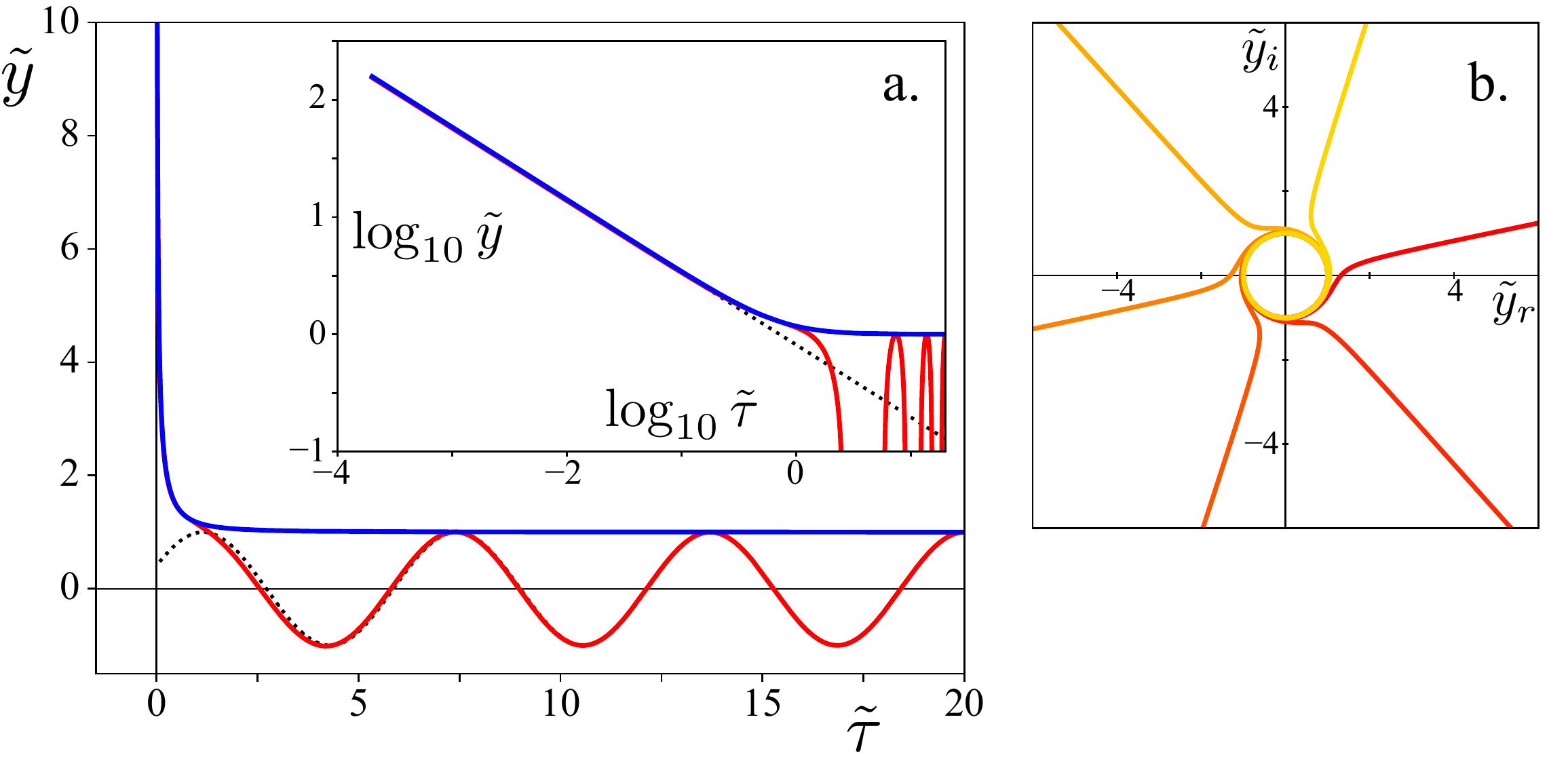}}% Images in 100% size
\caption{a. Solution of the amplitude equation for $B=1$. The main figure shows the amplitude $\tilde{y}$ as a function of time remaining to impact $\tilde{\tau}$. The red curve shows the real part of the solution $\tilde{y}_r(\tilde{\tau})$ with boundary conditions $\tilde{y}(20) = 1$,  $\dot{\tilde{y}}(20) = 0$ at $\tilde{\tau}_0=20$, whereas the blue curve shows the magnitude $|\tilde{y}(\tilde{\tau})|$ of the complex solution starting from the same boundary conditions. The dotted black line corresponds to the approximate solution in the large-$\tilde{\tau}$ limit (Eq.~\eqref{eq:earlylimeq}), again starting from the same point. The inset shows the same data in a doubly logarithmic plot and the dotted black line the approximate solution $\tilde{y} =\tilde{y}_-\tilde{\tau}^{\kappa_-} $ in the small-$\tilde{\tau}$ limit (Eq.~\eqref{eq:latelimeq}) with $\tilde{y}_- = 1.4$. b. Projection onto the complex plane $\tilde{y} = \tilde{y}_r + i\tilde{y}_i$ of solutions of the amplitude equation starting from the same boundary conditions as in a., but taken at different initial times $\tilde{\tau}_0 = 20 - n\pi/3$ for $n=0,...,5$. Clearly, after spending time on the unit circle (corresponding to oscillations), divergence may occur in any direction in the complex plane. The curve closest to the positive $\tilde{y}_r$-axis corresponds to the one plotted in figure a., i.e., $\tilde{\tau}_0 = 20$. 
}
\label{fig:yvstau}
\end{figure}

In Fig.~\ref{fig:yvstau}a we plot the real part $\tilde{y}_r$ of the dimensionless amplitude $\tilde{y}$ as a function of dimensionless time $\tilde{\tau}$ for $B=1$ (red curve). Clearly, for large $\tilde{\tau}$ the solution is oscillatory and is fitted nicely by the asymptotic solution \eqref{eq:earlylimsol} (dotted black curve). As expected, the solution starts to diverge when $\tilde{\tau} \approx \sqrt{B} = 1$, which may be better observed in the inset where the same data is plotted doubly logarithmically. Clearly, $\tilde{y}$ diverges as a power law, which is well fitted by the dotted black line, which corresponds to Eq.~\eqref{eq:latelimsol}, with $\tilde{y}_- = 1.4$. In the same plots, the blue curve corresponds to the magnitude $|\tilde{y}(\tilde{\tau})|$ of the complex solution with the same boundary conditions, which in this case happens to coincide with $\tilde{y}_r$. This is not always the case, since the divergence may happen in any direction in the complex plane, depending on initial conditions, as evidenced by Fig.~\ref{fig:yvstau}b.

\begin{figure}
 \centerline{\includegraphics[width=\textwidth]{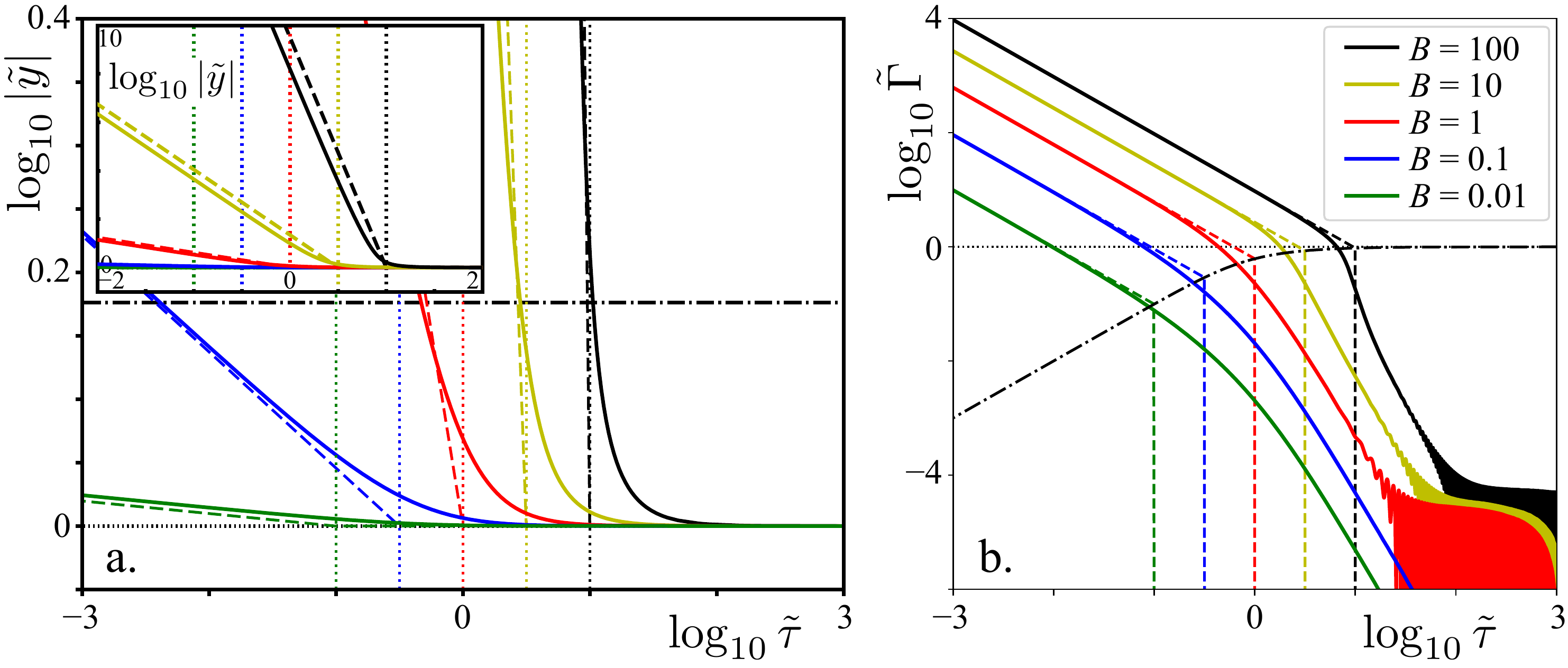}}% Images in 100% size
\caption{a. Doubly logarithmic plot of the absolute amplitude $|\tilde{y}|$ as a function of time $\tilde{\tau}$ for 5 different values of $B = 0.01, 0.1, 1, 10, 100$ (solid lines), in a close up near to the asymptotic early-time solution ($\tilde{\tau}>\sqrt{B}$). The dashed lines represent the approximate solution~\eqref{eq:approxsol}, whereas the horizontal black dashed-dotted line represents a threshold value $p=1.5$ (i.e., $\log_{10} p = 0.176$). Note that the differences between the full and approximate solutions are small for this value of $p$. The inset contains the same data plotted for a larger range in the vertical coordinate to indicate that all data diverge as a power law for $\tilde{\tau}\to 0$, with the strongly $B$-dependent exponent $\kappa_- = (\sqrt{4B+1}-1)/2$. The vertical dotted lines indicate the crossover point $\tilde{\tau}=\sqrt{B}$. b. Doubly logarithmic plot of the absolute amplitude growth rate $\tilde{\Gamma}$ (solid lines) vs. $\tilde{\tau}$. The dashed lines represent the approximate solution $\tilde{\Gamma}_\textrm{appr}$ and the dashed dotted black line corresponds to the onset value of $\tilde{\Gamma}_\textrm{appr}$ at $\tilde{\tau} = \sqrt{B}$.    
All quantities are dimensionless (see main text).}
\label{fig:absyGammavsB}
\end{figure}

For the purpose of our stability analysis it is sufficient to look at the magnitude $|\tilde{y}(\tilde{\tau})|$. Clearly, for $\tilde{\tau} \gg \sqrt{B}$, we find that 
 $|\tilde{y}(\tilde{\tau})| \approx 1$ (Eq.~\eqref{eq:earlylimsol}), whereas close to impact ($\tilde{\tau} \ll \sqrt{B}$) we have  $|\tilde{y}(\tilde{\tau})| \approx \tilde{y}_-\tilde{\tau}^{\kappa_-}$ (Eq.~\eqref{eq:latelimsol}). By simply 
%extrapolating both approximate solutions to 
matching both asymptotic solutions at the crossover point $\tilde{\tau} = \sqrt{B}$, we %can match the two solutions to 
obtain the unknown $\tilde{y}_-$ and arrive at a continuous approximate solution $|\tilde{y}|_\textrm{appr} \approx  |\tilde{y}|$
 \begin{equation}\label{eq:approxsol}
 \left|\tilde{y}\right|_\textrm{appr}(\tilde{\tau}) = \begin{dcases}
               \,\,\,\,\,\,1\qquad\qquad\qquad\qquad\textrm{for}\,\,\tilde{\tau}\geq\sqrt{B}\,,\\
               \left(\frac{\tilde{\tau}}{\sqrt{B}}\right)^{-\tfrac{1}{2}(\sqrt{4B+1}-1)}\qquad\,\,\,\textrm{for}\,\,\tilde{\tau}<\sqrt{B}\,.
            \end{dcases}
 \end{equation}
To understand how the solution depends on the parameter $B$, we plot the magnitude of the dimensionless amplitude $|\tilde{y}|$ (solid lines) together with the approximate solution $|\tilde{y}|_\textrm{appr}$ (dashed lines) for several different values of $B$ in Fig.~\ref{fig:absyGammavsB}a, all starting with $\tilde{y}(\tau_0) = 1$ and $\dot{\tilde{y}}(\tau_0) = 0$ at $\tilde{\tau}_0 = 1000$.  Clearly, all curves start to diverge for $\tilde{\tau}$ below $\sqrt{B}$, but the rate at which this happens strongly depends on $B$, consistent with the above approximate solution. Note that the approximate solutions only fairly coincide with the exact ones: Whereas the power-law exponent is highly accurate, one observes a shift between $|\tilde{y}|$ and $|\tilde{y}|_\textrm{appr}$ that is especially pronounced for large $B$ (inset of Fig.~\ref{fig:absyGammavsB}a). Here it is good to note that, although a matched asymptotic expansion can be constructed, it is algebraically complex and not converging quickly such that the approximate solution of Eq.~\eqref{eq:approxsol} is preferred. Also, the crossover point $\tilde{\tau}<\sqrt{B}$ marks exactly that point in time at which the solution can start to grow, i.e., where the time-dependent squared frequency becomes negative.
 
\subsection{Absolute amplitude growth rate}
\label{subsec:amplitudegrowthrate}

A second quantity that is worth looking at is the absolute amplitude growth rate $\tilde{\Gamma}$, which we define as the rate at which the absolute amplitude grows  
\begin{equation}\label{eq:defGamma}
\tilde{\Gamma} \equiv  - \frac{1}{|\tilde{y}|}\frac{d|\tilde{y}|}{d\tilde{\tau}}\,,
\end{equation}
where the minus sign guarantees that $\tilde{\Gamma} \geq 0$ in the divergent region ($\tilde{\tau} \downarrow 0$). This definition coincides with the growth rate as it is commonly defined in stability problems for the case of exponential growth. The absolute amplitude growth rate is plotted in Fig.~\ref{fig:absyGammavsB}b (solid lines) for the data found in Fig.~ ~\ref{fig:absyGammavsB}a, where one observes that for $\tilde{\tau} \gg \sqrt{B}$ the growth rate is very small, and subsequently rapidly increases for  $\tilde{\tau} \gtrsim \sqrt{B}$, until, for $\tilde{\tau}< \sqrt{B}$, the solution converges to a power law with exponent $-1$. This last feature is easily understood by computing the absolute amplitude growth rate from the approximate solution $|\tilde{y}|_\textrm{appr}$
\begin{equation}\label{eq:approxsolGamma}
\tilde{\Gamma}_\textrm{appr}(\tilde{\tau}) = -\frac{1}{|\tilde{y}|_\textrm{appr}}\frac{d|\tilde{y}|_\textrm{appr}}{d\tilde{\tau}} = 
\begin{dcases}
               \qquad\,\,0\qquad\qquad\qquad\,\,\textrm{for}\,\,\tilde{\tau}\geq\sqrt{B}\,,\\
               \frac{\sqrt{4B+1}-1}{2\tilde{\tau}}\qquad\qquad\textrm{for}\,\,\tilde{\tau}<\sqrt{B}\,,
            \end{dcases}
\end{equation}
which indeed corroborates the observed power-law exponent. In fact, the approximate absolute amplitude growth rate $\tilde{\Gamma}_\textrm{appr}$ (dashed lines) excellently coincides with the numerical solution $\tilde{\Gamma}$ in the late regime ($\tilde{\tau}<\sqrt{B}$), which once more confirms that the power-law exponent for $\tilde{y}$ in this limit is well predicted by the approximate solution. A final feature that is worth pointing to in Fig.~\ref{fig:absyGammavsB}b is that initially $\tilde{\Gamma}$ appears to oscillate with a small amplitude ($\sim 10^{-4}$) for $B\geq1$. Although this detail is not relevant for the stability question we are addressing, it is good to note that this is not some numerical inaccuracy, but a property of the equations, as is shown in Appendix~\ref{appB} where we will derive a set of equations for $\tilde{\Gamma}$, which is not only elegant, but also easier to solve numerically. 

\subsection{Determining stability criteria}
\label{subsec:amplitudestabcrit}

An important question we will need to address is to determine the onset time of the instability, i.e., the time at which the solution $|\tilde{y}(\tilde{\tau})|$ becomes unstable. We will now set out to define and determine this threshold value. 

From the above discussion it is clear that $|\tilde{y}(\tilde{\tau})|$ is stable at early times, and starts to diverge when $\tilde{\tau}$ becomes %sufficiently
smaller than $\sqrt{B}$. What is complicating this task, however, is that the divergence is not uniform, but strongly depends of the value of the parameter $B$. Naively, one could just set instability onset to the time below which the (approximate) dimensionless amplitude first starts to rise above $1$, i.e., at $\tilde{\tau} = \sqrt{B}$. This would however seriously over-predict the onset time for small values of $B$. E.g., looking at the blue curve corresponding to $B=0.1$ in Fig.~\ref{fig:absyGammavsB}a one observes that $\sqrt{B} \approx 0.316$, whereas the solution only surpasses the dashed-dotted black line, indicating a growth of the amplitude to just $1.5$ times its original size, at $\tilde{\tau} \approx 3.78\cdot10^{-3}$, which is almost two orders of magnitude smaller than $\sqrt{B}$. 

The need to distinguish these slow growing modes with small $B$, forces us %therefore choose 
to define the onset time by setting a threshold value $p$ for the absolute amplitude $|\tilde{y}|$ to determine the onset time and define it as:
\begin{equation}\label{eq:tauonsdef}
 |\tilde{y}|(\tilde{\tau}_\textrm{ons}) = p\,.
\end{equation}  
This does more justice to the dynamics of the amplitude equation, but comes at the expense of having to introduce an additional parameter to the problem, namely, the threshold value $p$. Also, setting a threshold value to the full, numerical solution $|\tilde{y}|$ will not lead to a practical algebraic expression. However, using the approximate solution~\eqref{eq:approxsol}, we can directly solve $|\tilde{y}|(\tilde{\tau}_\textrm{ons}) = p$ as
\begin{equation}\label{eq:tauons}
\tilde{\tau}_\textrm{ons, appr} = \sqrt{B} \,p^{-2/(\sqrt{4B+1}-1)} = \sqrt{B}\exp\left[-\frac{2\log(p)}{\sqrt{4B+1}-1}\right]\,.
\end{equation}
Coincidentally, at a reasonable threshold value $p=1.5$ (horizontal dashed-dotted black line in Fig.~\ref{fig:absyGammavsB}a) the discrepancy between $|\tilde{y}|$ and $|\tilde{y}|_\textrm{appr}$ is observed to be very small. Thus, for such threshold values we may expect the approximate onset times to lie close to the numerical ones.   

\begin{figure}
 \centerline{\includegraphics[width=0.7\textwidth]{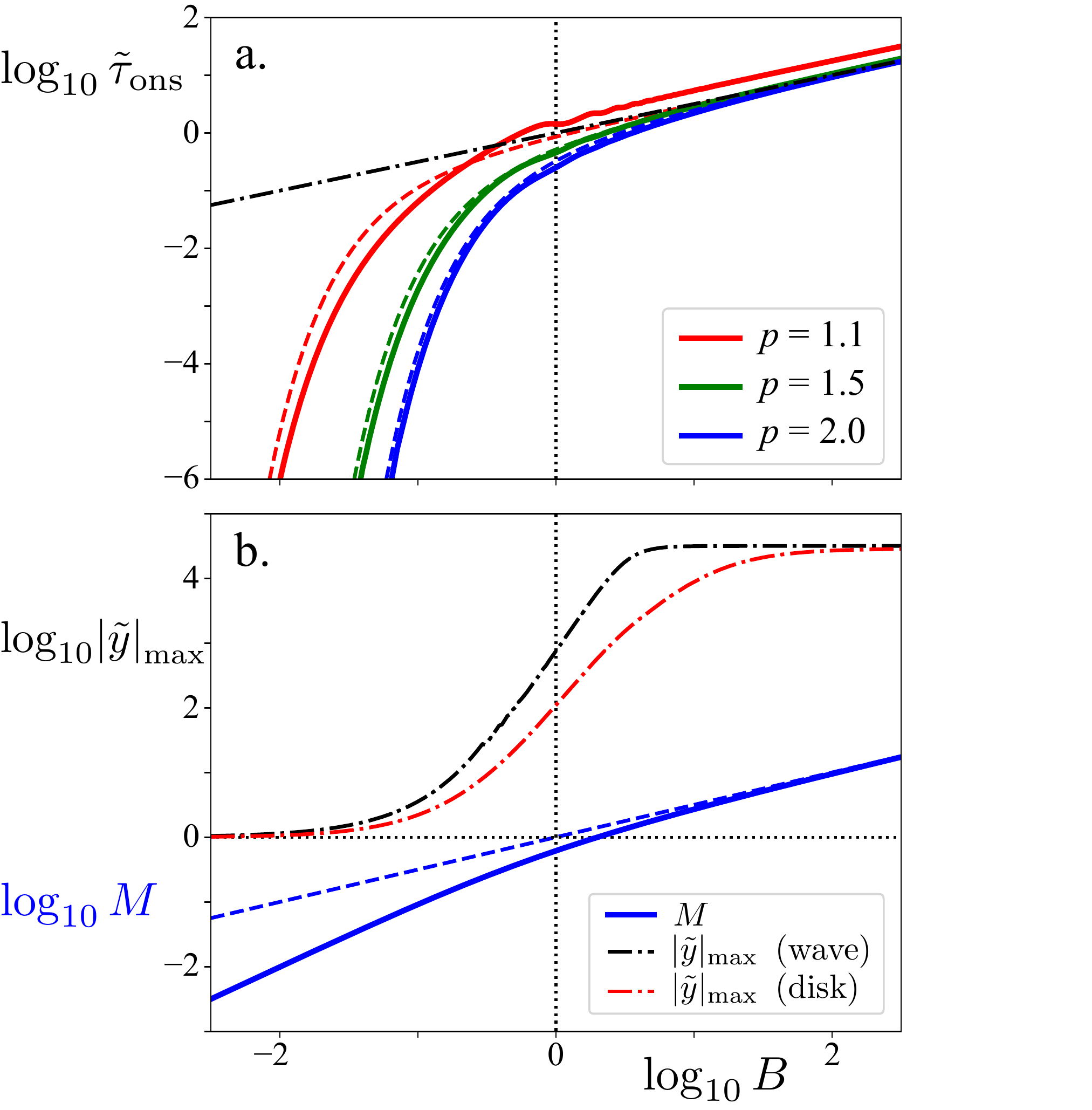}}% Images in 100% size
\caption{a. Onset time $\tilde{\tau}_\textrm{ons}$ as a function of $B$ for three different threshold values $p = 1.1$, $1.5$, $2.0$, both numerically from the direct solution $\tilde{y}(\tilde{\tau})$ (solid lines) and using the approximate expression \eqref{eq:tauons} (dashed lines). Note that the undulations that are visible, e.g., in the solution for $p=1.1$, are due to initial conditions, which are here taken to be $\tilde{y}(\tilde{\tau}_0) =1$ and $\dot{\tilde{y}}(\tilde{\tau}_0) =0$, for $\tilde{\tau}_0 = 4B$. The undulations may be suppressed taking larger values for $\tilde{\tau}_0$. The dashed dotted line represents the semi-classical result $\tilde{\tau}_\textrm{sclas} = \sqrt{B}$. b. Multiplication factor $M$ as a function of $B$. Again, the dashed-dotted line denotes the semi-classical result $M_\textrm{sclas} = \sqrt{B}$. Also plotted are the numerically determined maximum growth factors $|\tilde{y}|_\textrm{max}$ for two representative parameter settings corresponding to wave (black dashed-dotted line) and disk impact (red dashed-dotted line). See Appendix~\ref{appF} for details of this estimate.
}
\label{fig:TauonsVsB}
\end{figure}

In Fig.~\ref{fig:TauonsVsB}a we plot the numerically determined onset times $\tilde{\tau}_\textrm{ons}$ (solid lines) together with the approximation $\tilde{\tau}_\textrm{ons, appr}$ from the above equation (dashed lines) for different values of the threshold $p$. Clearly, the approximate solution gives an accurate description of the actual onset time, especially for intermediate threshold values, $p\approx 1.5$). For the smallest $p$ value plotted, there is a clear shift of the numerical result to slightly larger onset times. 

Equally significant is the observation that the onset time quantitatively depends on the choice of the threshold value $p$. This can be easily understood when looking back at Fig.~\ref{fig:absyGammavsB}a, where one observes that, especially for smallest values of $B$, the power-law exponent in the divergent region, $\kappa_- = \tfrac{1}{2}(1-\sqrt{4B+1})$, becomes vanishingly small, corresponding to large shifts in onset time for varying $p$. This is an issue that especially impacts this time-dependent analysis, but is also worthy of consideration for the classical analysis, as discussed in the next subsection.

Given that the onset time depends on the choice of the threshold value, it may be good to look for an additional quantity that may provide some insight in the speed with which the solution diverges. Naturally, one may be tempted to use the absolute amplitude growth rate $\tilde{\Gamma}$, but this quantity still depends on time in the divergent regime, and additionally is also affected by the non-dimensionalization of the time coordinate. One can however define another quantity that is not affected by the above two complications, namely the multiplication factor $M = \tilde{\Gamma}\tilde{\tau}$, which is simply the product of $\tilde{\Gamma}$ and the amount of time remaining until impact, $\tilde{\tau}$. Since asymptotically,  $\tilde{\Gamma}\sim1/\tilde{\tau}$, once in the diverging regime, $M$ is not expected to depend on time and, given the excellent agreement of the approximate and actual expressions in this regime, one may simply compute $M$ as          
\begin{equation}\label{eq:M}
M = \tilde{\Gamma}\,\tilde{\tau} \approx \frac{\sqrt{4B+1}-1}{2\tilde{\tau}}\,\tilde{\tau} = \tfrac{1}{2}(\sqrt{4B+1}-1)\qquad\qquad(\textrm{for}\,\,\tilde{\tau}\ll \sqrt{B})\,.
\end{equation}
This expression is plotted in Fig.~\ref{fig:TauonsVsB}b. With the definition of $\tilde{\Gamma}$ (Eq.~\eqref{eq:defGamma}), one may adopt $M$ as a measure for the relative increase $\Delta |\tilde{y}|/|\tilde{y}|$ in the diverging time interval $\tilde{\tau}_\textrm{ons}$, but of course cannot be identified with it, since the true amplitude is diverging with time $\tilde{\tau}$. This implies that, even within the linear context, $M$ will be a lower bound on the actual increase in amplitude, which may even be orders of magnitude larger than $M$. 

This is confirmed by numerically solving for the maximally possible amplitude that the solution can obtain, realizing that the amplitude $|y(\tau)|$ can never become larger than the remaining gap width ($U_0\tau$) between the solid and the free surface. This condition leads to a maximum growth factor $|\tilde{y}|_\textrm{max} = |y|_\textrm{max}/y_0$ of the original amplitude $y_0$, which are also plotted in Fig.~\ref{fig:TauonsVsB}b for some typical parameter values. More details on the calculation can be found in Appendix~\ref{appF} and in the next section, where the two curves will be discussed in some greater detail. It suffices to note that the more realistic estimate of growth $|\tilde{y}|_\textrm{max}$ is orders of magnitude larger than $M$, at the expense of explicitly depending on all relevant system parameters, including the initial amplitude $y_0$.

Finally, the divergence of the multiplication factor $M$ for large $B$ (i.e., small wavelength $\lambda$) may look puzzling at first sight, but one should realize that the amount of time in which this divergence takes place scales with $1/\sqrt{A} \sim \lambda^{3/2}$ and consequently becomes vanishingly small for small $\lambda$.

\subsection{Stability criteria and (semi-)classical stability analysis}
\label{subsec:stabcritclass}

The final subject to discuss in this Section is how all of the above relates to a traditional stability analysis, where in Eq.~\eqref{eq:nondimampleq} the factor $\tilde{\Omega}^2 \equiv 1-B/\tilde{\tau}^2$ would be time independent. Before looking at the conditions under which Eq.~\eqref{eq:nondimampleq} may behave classically, let us first naively neglect the time-dependence of the equations and write down the results that we may classically expect. This corresponds to assuming (wrongly, as we know) that $\tilde{\Omega}^2$ changes slowly with time, a situation which we will call this the \emph{semi-classical} picture. First of all, Eq.~\eqref{eq:nondimampleq} would become unstable as soon as $\tilde{\Omega}^2$ becomes zero, leading to the semi-classical onset time
\begin{equation}\label{eq:tauonssclass}
\tilde{\tau}_\textrm{ons,sclass} = \sqrt{B}\,,
\end{equation}
which result we have plotted in Fig.~\ref{fig:TauonsVsB}a as the black dashed-dotted line. The semi-classical absolute amplitude growth rate is simply zero above $\tilde{\tau}_\textrm{ons, sclass}$ and $\sqrt{-\tilde{\Omega}^2}$ below it, i.e.,
\begin{equation}\label{eq:Gammasclass}
\tilde{\Gamma}_\textrm{sclass}(\tilde{\tau}) = 
\begin{dcases}
               \quad\,\,\,0\qquad\qquad\qquad\qquad\,\,\textrm{for}\,\,\tilde{\tau}\geq\sqrt{B}\,,\\
               \frac{\sqrt{B-\tilde{\tau}^2}}{\tilde{\tau}} \approx  \frac{\sqrt{B}}{\tilde{\tau}}\qquad\qquad\textrm{for}\,\,\tilde{\tau}<\sqrt{B}\,,
            \end{dcases}
\end{equation} 
where the last approximate equality holds at the very last stages ($\tilde{\tau} \ll \sqrt{B}$). Clearly, using this last expression we may write down a semi-classical multiplication factor as
\begin{equation}\label{eq:Msclass}
M_\textrm{sclass} = \tilde{\Gamma}_\textrm{sclass}\,\tilde{\tau} = \sqrt{B-\tilde{\tau}^2} \approx \sqrt{B}\,.
\end{equation}
It is this expression that we have plotted together with the full expression~\eqref{eq:M} in Fig.~\ref{fig:TauonsVsB}b. It needs to be stressed that this semi-classical picture does \emph{not} represent the result of traditional time-independent stability analysis, but what one would get if one would naively apply its results to this time-dependent case. Based on the above one could say that the full expressions converge to the semi-classical ones in the limit $B \gg 1$, and drastically diverge when $B \lesssim 1$. The latter is therefore the region where one would expect to obtain results that strongly diverge from the time-independent picture. Most specifically, for small $B$ solutions are seen to be considerably stabilized with respect to the time-independent case: the onset time is strongly suppressed (which implies that they become unstable at a much later point in time, i.e., at smaller $\tilde{\tau}$) and also the multiplication factor is much smaller ($\sim B$ as compared to $\sim \sqrt{B}$ in the semi-classical case).

Finally, we ask ourselves the question under what conditions taking a time-independent approach would be justified. This is expected to be the case when $\tilde{\Omega}^2$ changes sufficiently slowly as a function of time, such that the dynamics is only minorly influenced. This implies that the time derivative of $\tilde{\Omega}^2$ must be sufficiently small in the region where the solution becomes unstable, i.e., close to onset time:
\begin{equation}\label{eq:classcond}
\left.\frac{d(\tilde{\Omega}^2)}{d\tilde{\tau}}\right|_{\tilde{\tau}_\textrm{ons}} = \left.\frac{2B}{\tilde{\tau}^3}\right|_{\tilde{\tau} = \sqrt{B}}=\frac{2}{\sqrt{B}}\ll1\qquad\Rightarrow\qquad B\gg4\,,
\end{equation}
confirming that a time-independent approach may be justified for very large values of $B$.

\section{Stability diagrams}
\label{sec:stability}

Now that we have analyzed the solutions of the amplitude equation \eqref{eq:nondimampleq} and its properties, we turn to its applications for the two cases that we have introduced in Section~\ref{sec:problem}. The basis for this Section will be the dimensional form of the approximate onset time solution~\eqref{eq:tauons}, namely 
\begin{equation}\label{tauons_dim}
\tau_\textrm{ons} = \sqrt{\frac{B}{A}}\exp\left[-\frac{2\log(p)}{\sqrt{4B+1}-1}\right]\,,
\end{equation}
where we used that $\tilde{\tau}=\tau\sqrt{A}$, together with the multiplication factor $M$ (Eq.~\eqref{eq:M}). 
Now, since both $A$ and $B$ are functions of the wave number $k$, so will the onset time $\tau_\textrm{ons}$, and the equivalent procedure of finding a marginally stable wavelength in this time-dependent case is to look for the wave number that becomes unstable first, i.e., to look for which value of $k$ the function $\tau_\textrm{ons}(k)$ attains a maximum.    

\begin{figure}
\centerline{\includegraphics[width=0.7\textwidth]{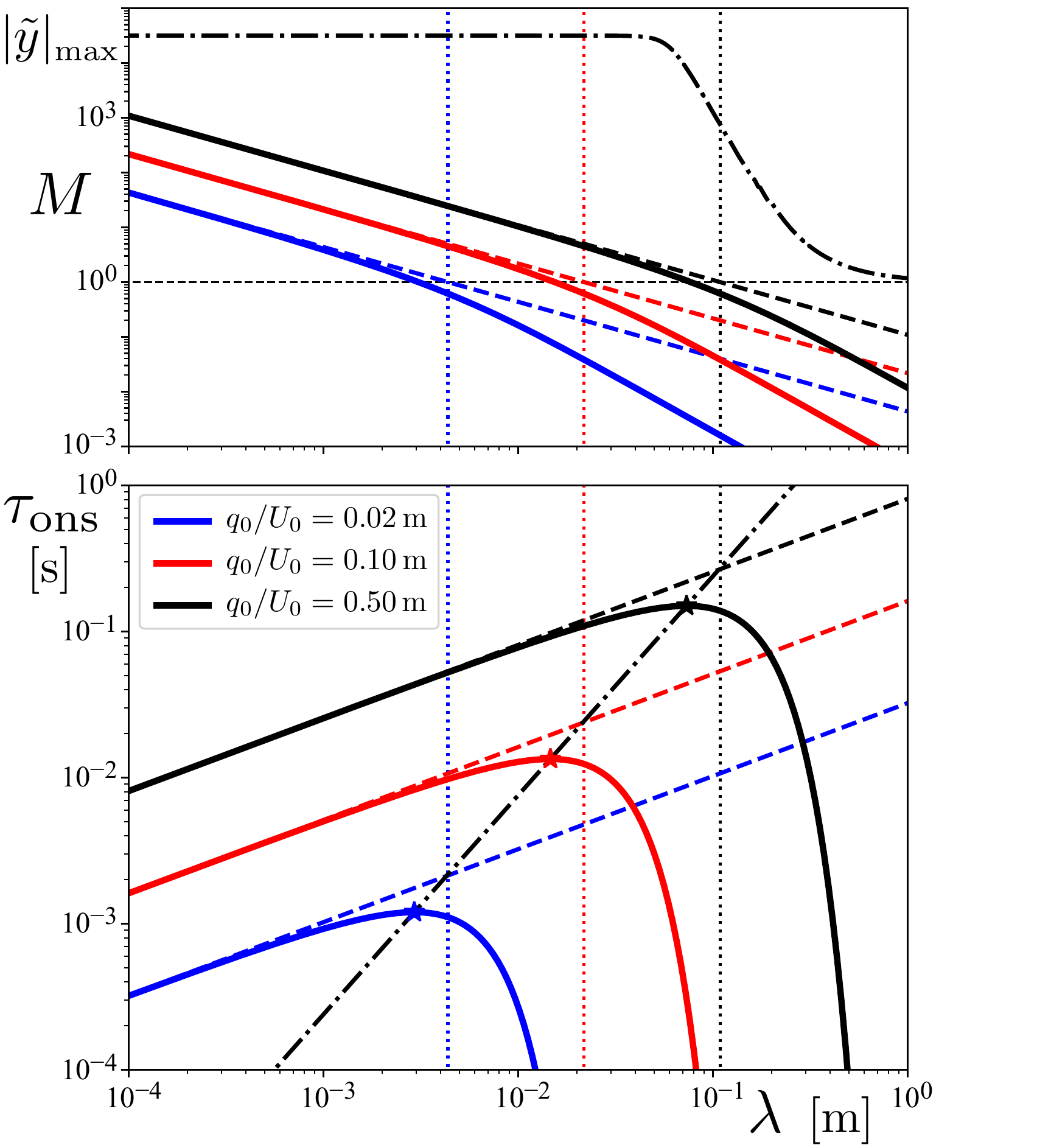}}% Images in 100% size
\caption{Stability diagram for a wave impacting a vertical wall. \emph{Bottom:} Doubly logarithmic plot of the onset time $\tau_\textrm{ons}$ versus wavelength $\lambda$ for three different values of the gas flow rate parameter $q_0/U_0$ (solid lines; see legend), with $p=1.5$. The dashed lines represent the semi-classical result $\tau_\textrm{ons,sclass}$ and vertical dotted lines indicate for which $\lambda$ the value $B=1$ is reached. The dashed-dotted black curve indicates the location of the marginal onset time $\tau_\textrm{marg}$ as a function of the marginal wavelength $\lambda_\textrm{marg}$. \emph{Top:} Multiplication factor $M$ as a function of the wavelength $\lambda$ for the same three cases as in the bottom plot (solid lines), together with the semiclassical result $M_\textrm{sclass}$ (dashed lines). The dashed-dotted black line corresponds to the numerically calculated maximum growth factor $|\tilde{y}|_\textrm{max}$ (see Appendix~\ref{appF}) corresponding to the largest gas flow rate parameter $q_0/U_0 = 0.50$ m and an impact velocity $U_0 = 6.0$ m/s.
Densities and interfacial tension correspond to those of water and air at $20$ $^\circ$C and atmospheric pressure ($\rho_l = 998$ kg/m$^3$, $\rho_g = 1.20$ kg/m$^3$, and $\sigma = 0.073$ N/m).
}
\label{fig:WaveImpact}
\end{figure}
 
\subsection{Wave impact against a vertical wall} 
\label{subsec:waveimpact}
 
Starting with the case where a wave impacts a vertical wall, we first notice that since gravity acts parallel to the wall, we may set $g=0$ in the expression \eqref{eq:Adef} for $A$, such that $A = \alpha_3k^3$ and $B=\beta_2k^2$, with which
\begin{eqnarray}\label{tauons_wave}
\tau_\textrm{ons,w}(k) &=& \sqrt{\frac{\beta_2}{\alpha_3}}k^{-1/2}\exp\left[-\frac{2\log(p)}{\sqrt{4\beta_2k^2+1}-1}\right]\nonumber\\
&=& \sqrt{w(\delta)(1+\delta)\frac{\rho_l}{\sigma}}\,\frac{q_0}{U_0}k^{-1/2}\exp\left[-\frac{2\log(p)}{\sqrt{4\,w(\delta)\frac{q_0^2}{U_0^2}k^2+1}-1}\right]\,,
\end{eqnarray}
where in the last equation we have inserted the expressions \eqref{eq:Adef} and \eqref{eq:Bdef} for $\alpha_3$ and $\beta_2$ respectively, and where we have defined the function $w(\delta)$ as
\begin{equation}\label{w_delta}
w(\delta) = \frac{\delta}{(1+\delta)^2} \,
\end{equation}
for notational convenience. Note that in the limit $\delta \ll 1$, which is generally satisfied, $w(\delta) \approx \delta$.
%Note that it is mainly the elaborate functions of the density ratio $\delta$ which makes this expression look complex, and that both these expressions may be replaced by $\delta$ itself in the limit . 
  
To plot the expressions for the onset time and multiplication factor, we take the air-water interface at $20$ $^\circ$C and standard atmospheric pressure as an example. This fixes the the liquid and gas densities and the interfacial tension to $\rho_l = 998$ kg/m$^3$, $\rho_g = 1.20$ kg/m$^3$, and $\sigma = 0.073$ N/m. In Fig.~\ref{fig:WaveImpact} we plot the onset time $\tau_\textrm{ons}$ as a function of the wavelength $\lambda = 2\pi/k$ for different values of $q_0/U_0$, together with the semi-classical result
\begin{equation}\label{tauonssclass_wave}
\tau_\textrm{ons,w,sclass}(k) = \sqrt{\frac{B}{A}} = \sqrt{\frac{\beta_2}{\alpha_3}}\frac{1}{\sqrt{k}}= \sqrt{w(\delta)(1+\delta)\frac{\rho_l}{\sigma}}\,\frac{q_0}{U_0}\frac{1}{\sqrt{k}}\,.
\end{equation}
Clearly, whereas the semi-classical result $\tau_\textrm{ons,w,sclass}(\lambda)$ diverges for $\lambda\to\infty$, indicating that long wavelengths are always unstable, the time-dependent result $\tau_\textrm{ons,w}(\lambda)$ exhibits a maximum at some finite, marginally stable wavelength $\lambda_\textrm{marg}$, for which the liquid interface first becomes unstable as the wave approaches the wall (i.e., for $\tau\to 0$). This marginal wavelength is thus solely selected as a consequence of the time dependence of the process. 
 
To find the marginal wavelength we need to determine the location of the maximum by solving $d\tau_\textrm{ons,w}/dk = 0$, which is done in Appendix~\ref{appD} and results in
\begin{equation}\label{eq:lambdamarg}
\lambda_\textrm{marg} = 2\pi \frac{m(\delta)}{g(p)}\,\frac{q_0}{U_0}\,,
\end{equation}
where the functions $g(p)$ and $m(\delta)$ are defined in Eqs.~\eqref{eqD:g_pfinal} and \eqref{eqD:m_delta}. Here, note that $m(\delta) \approx \sqrt{\delta}$ in the case that $\delta \ll 1$. Reinserting the expression for $k_\textrm{marg} = 2\pi/\lambda_\textrm{marg}$ into Eq.~\ref{tauons_wave}, provides us with an expression for the marginal onset time $\tau_\textrm{marg} = \tau_\textrm{ons,w}(k_\textrm{marg})$  
\begin{equation}\label{eq:taumarg}
\tau_\textrm{marg} = h(p)\,n(\delta)\,\left(\frac{\rho_l}{\sigma}\right)^{1/2}\left(\frac{q_0}{U_0}\right)^{3/2}\,,
\end{equation}
where $h(p)$ and $n(\delta)$ are defined in Eqs.~\eqref{eqD:h_pfinal} and \eqref{eqD:n_delta}. The line connecting the different $(\lambda_\textrm{marg},\tau_\textrm{marg})$-pairs for varying gas flow rate parameter $q_0/U_0$ is plotted as the dashed-dotted line in Fig.~\ref{fig:WaveImpact}, through the marginal wavelengths and onset times of the three plotted curves, which are indicated by the asterisks. 

From the above equations it follows that the marginal wavelength only depends on the gas-flow rate parameter $q_0/U_0$, to which it is proportional, and on the density ratio, through $\lambda_\textrm{marg} \sim \delta^{1/2}(q_0/U_0)$ using the small $\delta$ approximation provided in Eq.~\eqref{eqD:m_delta}. There is no surface tension dependence. In contrast, the onset time does depend on both surface tension and gas flow rate, namely as $\tau_\textrm{marg} \sim  \delta^{3/4}(\rho_l/\sigma)^{1/2}(q_0/U_0)^{3/2}$, now using the small $\delta$ approximation for $n(\delta)$ (Eq.~\eqref{eqD:n_delta}). As a result, the gas velocity at the marginal onset time, $U_{g,\textrm{marg}} = U_g(\tau_\textrm{marg})$, can be estimated as $U_{g,\textrm{marg}} \sim  \delta^{-3/4}(\rho_l/\sigma)^{-1/2}(q_0/U_0)^{-1/2}$, which implies that the marginal gas speed decreases with increasing $q_0/U_0$.

The multiplication factor $M$ decreases with increasing wavelength, which can be traced back directly to the fact that $B \sim k^2$. For the threshold value $p = 1.5$ for which the stability diagram has been plotted $M$ is larger than unity for wavelengths smaller than $\lambda_\textrm{marg}$ but above this marginal value rapidly decreases to negligibly small values. This is consistent with the sharp cutoff of the onset time with increasing $\lambda$ where $\tau_\textrm{ons}$ drops by over five orders of magnitude within a decade of $\lambda$.

Turning to the numerically determined maximum growth factor $|\tilde{y}|_\textrm{max}$ corresponding to the largest gas flow rate parameter $q_0/U_0 = 0.50$ m (dashed-dotted black line in the top figure~\ref{fig:WaveImpact}) we see that at the marginal onset time $\tau_\textrm{marg}$ (black star) the initial amplitude will grow by a factor $|\tilde{y}|_\textrm{max} \approx 1\cdot10^5$ ($\approx 2.5$ mm for an initial amplitude of $25$ nm). This confirms that the growth rate of the marginal wavelength is sufficient to be observable. Also note that for small wavelengths (i.e., where $M$ diverges) $|\tilde{y}|_\textrm{max}$ tends to a constant large value, implying that all small wavelength disturbances may grow approximately by the same amount. It is good to realize that $\lambda_\textrm{marg}$ is the first disturbance to start growing, whereas the other modes may only start to grow later (i.e., at smaller values of $\tau$), when the surface has already been deformed with wavelength $\lambda_\textrm{marg}$.   
 
\subsection{Disk impact on a liquid surface} 
\label{subsec:diskimpact}
 
 \begin{figure}
\centerline{\includegraphics[width=0.7\textwidth]{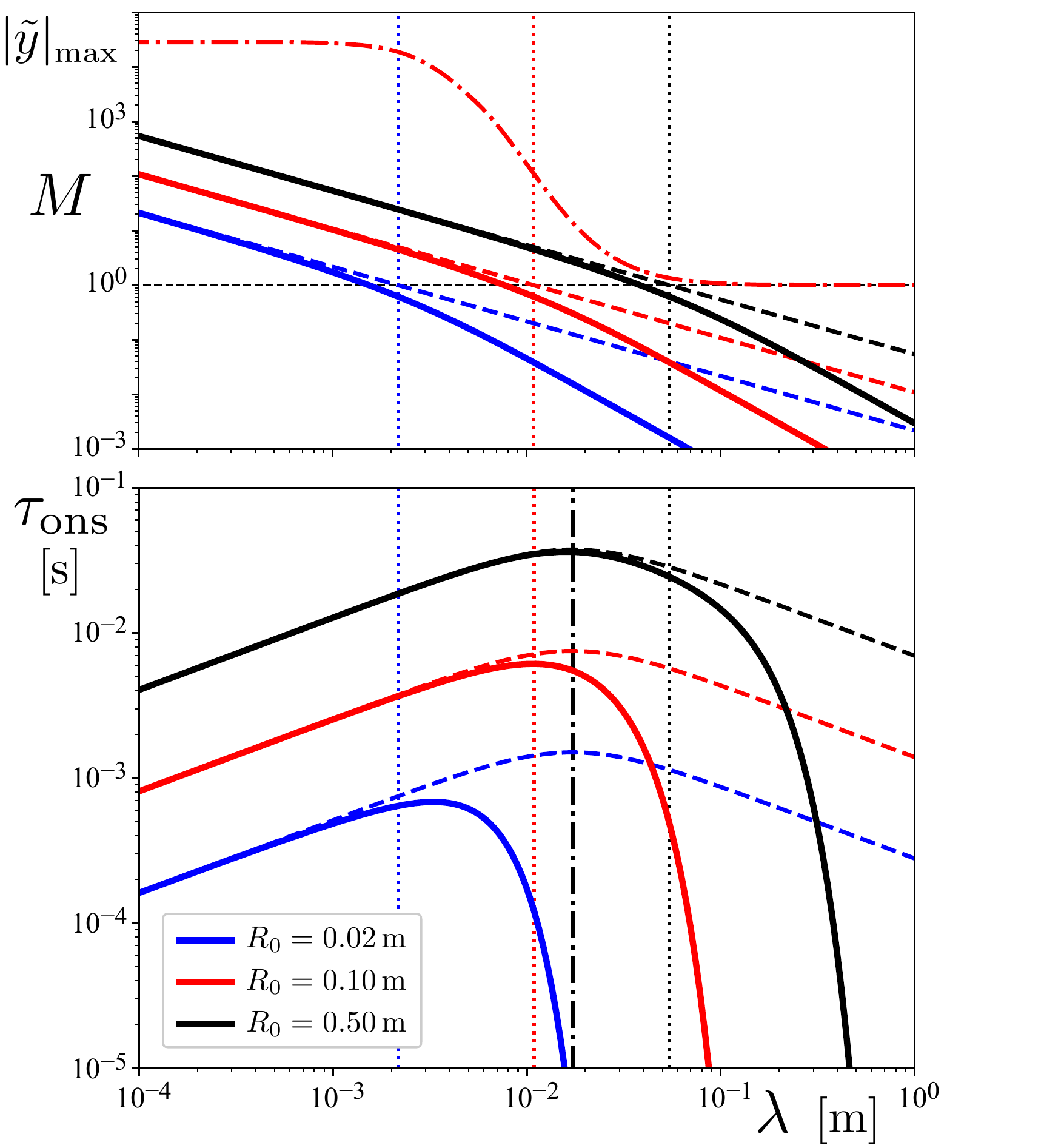}}
\caption{Stability diagram for a disk impacting onto a liquid surface. \emph{Bottom:} Doubly logarithmic plot of the onset time $\tau_\textrm{ons}$ versus wavelength $\lambda$ for three different values of the disk radius $R_0 = 2q_0/U_0$ (solid lines; see legend), with $p=1.1$. The dashed lines represent the semi-classical result $\tau_\textrm{ons,sclass}$ and vertical dotted lines indicate for which $\lambda$ the value $B=1$ is reached. The vertical dashed-dotted black curve indicates the location of the semi-classical marginal wavelength $\lambda_\textrm{marg,sclass}$. \emph{Top:} Multiplication factor $M$ as a function of the wavelength $\lambda$ for the same three cases as in the bottom plot (solid lines), together with the semiclassical result $M_\textrm{sclass}$ (dashed lines). The dashed-dotted red line corresponds to the numerically calculated maximum growth factor $|\tilde{y}|_\textrm{max}$ (see Appendix~\ref{appF}) corresponding to the middle radius $R_0 = 0.10$ m and an impact velocity $U_0 = 2.0$ m/s. Densities and interfacial tension correspond to those of water and air at $20$ $^\circ$C and atmospheric pressure ($\rho_l = 998$ kg/m$^3$, $\rho_g = 1.20$ kg/m$^3$, and $\sigma = 0.073$ N/m).
}
\label{fig:DiskImpact}
\end{figure}

We subsequently turn to the  impact of a disk on a liquid surface, in which case the gravitational term needs to be included in the equations. This implies that $A = \alpha_1k + \alpha_3k^3$ and $B=\beta_2k^2$, with which
\begin{eqnarray}\label{tauons_disk}
&&\tau_\textrm{ons,d}(k) = \sqrt{\frac{\beta_2k^2}{\alpha_1k + \alpha_3k^3}}\exp\left[-\frac{2\log(p)}{\sqrt{4\beta_2k^2+1}-1}\right]\nonumber\\
&=& \sqrt{\frac{\delta}{1+\delta}}\,\frac{R_0k}{2\sqrt{(1-\delta)gk + \frac{\sigma}{\rho_l}k^3}}\exp\left[-\frac{2\log(p)}{\sqrt{w(\delta)R_0^2k^2+1}-1}\right]\,,
\end{eqnarray}
again using \eqref{eq:Adef} and \eqref{eq:Bdef} for $\alpha_1$, $\alpha_3$ and $\beta_2$ to arrive at the last expression, and identifying $q_0/U_0 = R_0/2$.  
 
Again we use the example of an air-water interface at $20$ $^\circ$C and standard atmospheric pressure to plot the onset time as a function of wavelength in Fig.~\ref{fig:DiskImpact}. It is compared to the semi-classical result which in this case equals
\begin{equation}\label{tauonssclass_disk}
\tau_\textrm{ons,d,sclass}(k) = \sqrt{\frac{B}{A}} = \sqrt{\frac{\beta_2k^2}{\alpha_1k + \alpha_3k^3}}
=\tfrac{1}{2} \sqrt{\frac{\delta}{1+\delta}}\,\sqrt{\frac{\rho_l}{(1-\delta)\rho_lgk + \sigma k^2}}R_0 k\,.
\end{equation}
For disk impact, the semi-classical result is non-monotonic (as can be seen in Fig.~\ref{fig:DiskImpact} where it is plotted for three different values of the disk radius $R_0$) and attains a maximum for the marginal wavelength, which is calculated from the condition $d\tau_\textrm{ons}/dk = 0$, which leads directly to 
\begin{equation}\label{lambdamargsclass_disk}
\lambda_\textrm{marg,sclass} = 2\pi\sqrt{\frac{\alpha_3}{\alpha_1}} =   2\pi\sqrt{\frac{\sigma}{(1-\delta)\rho_l g}}\,. 
\end{equation} 
This is $2\pi$ times the capillary length, and as a consequence, independent of $q_0/U_0=R_0/2$. As may be expected from this fact, it is indeed exactly the same result as one obtains in a classical Kelvin-Helmholtz stability analysis, i.e., with a constant gas flow over the interface. The corresponding semi-classical marginal onset time is obtained by inserting $k_\textrm{marg,sclass}$ into Eq.~\eqref{tauonssclass_disk}
\begin{equation}\label{taumargsclass_disk}
\tau_\textrm{marg,sclass} = \frac{\beta_2^{1/2}}{(4\alpha_1\alpha_3)^{1/4}} = \tfrac{1}{4}\sqrt{\frac{2\delta}{1+\delta}}\left(\frac{\rho_l}{(1-\delta)\sigma g}\right)^{1/4}R_0\,. 
\end{equation} 
In the limit of small $\delta$ we have $\tau_\textrm{marg,sclass} \sim \delta^{1/2}(\rho_l/(\sigma g))^{1/4}R_0$. Clearly, also in this disk-impact case the onset time grows with $q_0/U_0 = R_0/2$, but not as fast as in the wave impact case, where we don't have a proportionality, but a power law with exponent $3/2$ (Eq.~\eqref{eq:taumarg}). We again estimate the gas velocity at this semi-classical marginal onset time, $U_{g,\textrm{marg,sclass}} = U_g(\tau_\textrm{marg,sclass})$, can be estimated as $U_{g,\textrm{marg,sclass}} \sim  \delta^{-1/2}(\sigma g/\rho_l)^{1/4}$, which just as in the classical case is found to be independent of $R_0$.

Above we are discussing the semi-classical limit, which is plotted as the dashed lines in Fig.~\ref{fig:DiskImpact}. When we compare this to the full solution (solid lines), we see that for large values of $R_0$ the semi-classical results for the marginal wavelength and marginal onset time are a good approximation to the true ones, with a cut-off happening for larger wavelengths, but as soon as $R_0$ comes in the centimetre range, the cut-off starts to interfere with the maximum in the semi-classical result (represented by the dashed dotted black line in Fig.~\ref{fig:DiskImpact}), and the maximum shifts to smaller wavelengths, as can be clearly seen in Fig.~\ref{fig:DiskImpact} in the curve corresponding to $R_0 = 2.0$ cm. %In fact, a closer look at the value   

It is not easy to find a general analytic expression for the marginal wavelength in the case of the disk as it involves solving a fourth order polynomial equation which can be shown (Appendix~\ref{appD}) to lead to a function of the form 
\begin{equation}\label{eq:lambdamargdisk}
\lambda_\textrm{marg,d} = \sqrt{w(\delta)}\,R_0\,F\!\left(\log(p),\frac{\alpha_1\beta_2}{\alpha_3}\right) = \sqrt{w(\delta)}\,R_0\,F\!\left(\log(p),\frac{w(\delta)\,(1-\delta) \rho_l g R_0^2}{4\sigma}\right)\,,
%the missing factor 2\pi is of course absorbed in F()
\end{equation}
where $F(\log(p),\zeta) \sim \zeta^{-1/2}$ for large $\zeta$ (i.e., large $R_0$) in order for the function to have the correct asymptotics leading to Eq.~\eqref{lambdamargsclass_disk}.

Also in this case, the multiplication factor $M$ decreases with increasing wavelength, which is again connected to $B \sim k^2$, and is even identical to the wave impact case (identifying $q_0/U_0 = R_0/2$). We observe that the multiplication factor $M$ at the location of the maximum in the $\tau_\textrm{ons}$-curve for the lowest value of $R_0$ is significantly smaller for the disk impact (Fig.~\ref{fig:DiskImpact}) when comparing to wave impact. This can only partly be traced back to the somewhat smaller value of the threshold ($p=1.1$) used to produce the $\tau_\textrm{ons}(\lambda)$ and suggests that due to the reduced growth rate the instability may be suppressed altogether at these smaller disk sizes. 

In this case, the numerically determined maximum growth factor $|\tilde{y}|_\textrm{max}$ is chosen to correspond to the middle disk size $q_0/U_0 = R_0/2 = 0.050$ m (dashed-dotted red line in the top figure~\ref{fig:DiskImpact}). The initial amplitude will grow by a factor $|\tilde{y}|_\textrm{max} \approx 1\cdot10^2$ at the marginal onset time, which is much less than in the wave impact case and makes observation of the marginal wavelength questionable, leading to the tentative conclusion that the influence of including time-dependence for small disk sizes is not only to shift the marginal wavelength to smaller values and to retard the marginal onset time (i.e., to smaller $\tau_\textrm{marg}$), but also to suppress the growth rate of the marginally stable wavelength.  

\subsection{The role of the threshold value $p$} 
\label{subsec:rolep}

In all of the above analysis, there is a pronounced influence of the choice of the threshold value $p$ on the results presented in this Section. Generally speaking, for larger values of the threshold the marginal wavelength becomes smaller. For values of $p$ between $1.1$ and $2.0$ the function $g(p)$ may be approximated accurately by $g(p) \approx (4.6(p-1))^{1/2}$, which implies a factor $3$ change of the marginal wavelength when varying $p$ in that range. Clearly, using values smaller than $p=1.1$ makes little sense, both because the approximation becomes less accurate and a small threshold value is hardly predictive for instability to occur, especially for small values of $B$. On the other limit, taking values larger than $p=2.0$ one may argue that when the amplitude has grown to two times its original size, the instability onset must have occurred in the past. From a comparison of the exact and the approximate solutions to the amplitude equation (cf. Figs.~\ref{fig:absyGammavsB}a and \ref{fig:TauonsVsB}), one may infer that taking $p=1.5$ in general is a good compromise with sufficient accuracy to have predictive value.

Although the somewhat problematic task of having to choose a threshold value appears to be a feature of this particular time-dependent problem and without doubt is closely related to the fact that the equation parameters themselves are rapidly changing functions of time, it is good to realize that this problem is not completely absent in a classical stability analysis: When one obtains a marginally stable wavelength, then by definition its growth rate is exactly zero for the corresponding setting of the system parameters. So, it is a priori unclear by how much one needs to detune the system parameters to actually observe the instability in an experiment, which may be related to the amount of time one has to observe the instability (especially true for convective instabilities), or the extent to which one is able to control fluctuations in the undisturbed setup. A classical (although very non-linear) example that comes to mind is the onset of turbulence in pipe flow.

\subsection{Advective terms} 
\label{subsec:advectiveterms}

Finally, we take two steps back in our analysis. The first step back is to where we needed to introduce a moving reference frame in order to have a valid base state for our analysis, which was moving with a velocity of magnitude $U_l(\tau)$ in the negative $x$-direction (Eq.~\eqref{eq:liqspeedcond}). This implies that we now need to transform the result of our analysis back into the lab frame, namely by propagating it with a velocity $U_l(\tau)$ in the positive $x$-direction. 

The second step back goes to the point where in order to obtain the simple form of the amplitude equation \eqref{eq:governing2}, a transformation of the original time-dependent amplitude $\varepsilon(\tau)$ into a new amplitude $\bar{\varepsilon}(\tau)$ was introduced by means of a phase factor. To obtain the original amplitude we need to transform our result $\bar{\varepsilon}(\tau) (= y(\tau))$ back
\begin{equation}\label{eq:phasefactor}
\varepsilon(\tau) = \bar{\varepsilon}(\tau)\exp[-ikH(\tau)]\,,
\end{equation}
with $H(\tau)$ as defined in Eq.~\eqref{eq:modified}. Using $U_g(\tau) = q_0/(U_0\tau)$ we can compute this function as 
\begin{equation}\label{eq:H}
H(\tau) = \frac{2\delta}{1-\delta^2}\int_{-\infty}^t U_g(\tau')dt' = - \frac{2\delta}{1-\delta^2}\frac{q_0}{U_0}\left[\log(\tau) - \log(\infty)\right]\,.
\end{equation}
Although strictly speaking the term $\log(\infty)$ is divergent, we may fix it at some large value $\log(\tau_s)$ to verify that it only contributes some arbitrary constant phase factor that is immaterial for the result we are after, and further neglect it. Inserting the above expression in Eq.~\eqref{eq:phasefactor}, we find that it represents an oscillatory factor, with a frequency and wave speed that diverge as $\tau$ approaches zero, or quantitatively 
\begin{equation}\label{eq:wavespeed}
c_H(\tau) = \frac{\omega(\tau)}{k} = \frac{H(\tau)}{\tau} = \frac{2\delta}{1-\delta^2}\frac{q_0}{U_0}\frac{\log(\tau)}{\tau}\,.
\end{equation}
Now, we have identified two typical velocities that transform any unstable wave pattern, it will be advected with a velocity $U_l(\tau)$ and modulated with a wave speed $c_H(\tau)$. To estimate their magnitude we will now compare these speeds with the other reference speeds that are at our disposal, namely the gas speed $U_g(\tau)$ and the approach speed of the impacting wave or disk $U_0$, thereby focussing on the marginal onset time $\tau_\textrm{marg}$.

Starting with $U_l(\tau)$, we find that its ratio with the gas velocity is constant
\begin{equation}\label{eq:liquidtogasvelocity}
\frac{U_l(\tau)}{U_g(\tau)} = \frac{\delta}{1-\delta} \approx \delta \qquad(\textrm{for}\,\,\delta\ll1)\,,
\end{equation}
and small, since the gas to liquid density ratio $\delta$ is small. For $c_H(\tau)$ we find similarly
\begin{equation}\label{eq:wavetogasvelocity}
\frac{c_H(\tau)}{U_g(\tau)} = \frac{2\delta}{1-\delta^2}\log(\tau) \approx 2\delta \log(\tau_\textrm{marg}) \qquad(\textrm{for}\,\,\delta\ll1\,\,\textrm{and}\,\,\tau=\tau_\textrm{marg})\,.
\end{equation}
Clearly, also in this case, since the dependence on $\tau_\textrm{marg}$ is logarithmic, the density ratio dominates the result, and we may conclude that generally the intrinsic velocities that disturb the wave pattern are one to three orders of magnitude smaller than the velocity that creates it, the gas velocity$U_g(\tau)$. 

It is however good to note that, whereas $U_l(\tau)$ advects the pattern in the same direction as the gas velocity, the wave with speed $c_H(\tau)$ propagates in the opposite direction, since $\log(\tau_\textrm{marg}) < 0$.
Nevertheless, all are diverging for $\tau\to 0$. Therefore it is appropriate to also compare them with the velocity of approach, $U_0$, at the time of marginal onset. First we turn to the wave impact case, and compare the advective velocity $U_l(\tau)$ to $U_0$, for which
\begin{equation}\label{eq:liquidtoU0}
\frac{U_l(\tau_\textrm{marg})}{U_0} = \frac{\delta}{(1-\delta)n(\delta)h(p)}\sqrt{\frac{\sigma}{\rho_lU_0q_0}} \approx \delta^{1/4}\sqrt{\frac{\sigma}{\rho_lU_0q_0}}  \qquad(\textrm{for}\,\,\delta\ll1)\,,
\end{equation}
where we have used expression~\eqref{eq:taumarg} and its small $\delta$ approximation, and for simplicity took $h(p) \approx 1$. Note that $\delta^{1/4}$ is not very small ($0.2-0.3$), but that the argument of the square root is to be interpreted as an inverse Weber number ($\textrm{We} = \rho_lU_0q_0/\sigma$), where for all but the very smallest impacts, $\textrm{We}$ is of the order $100$ or more commonly even larger. That is, usually, the advection velocity $U_l(\tau_\textrm{marg})$ is at least one to two orders of magnitude smaller than the impact velocity $U_0$. A similar conclusion may be drawn for the wave speed $c_H(\tau)$ for which
\begin{equation}\label{eq:wavetoU0}
\frac{c_H(\tau_\textrm{marg})}{U_0} = \frac{2\delta\,\log(\tau_\textrm{marg})}{(1-\delta^2)n(\delta)h(p)}\sqrt{\frac{\sigma}{\rho_lU_0q_0}}  \approx 2\delta^{1/4} \log(\tau_\textrm{marg})\sqrt{\frac{\sigma}{\rho_lU_0q_0}} \qquad(\textrm{for}\,\,\delta\ll1)\,,
\end{equation}
which is however a bit larger than the previous ratio due to the factor $2 \log(\tau_\textrm{marg})$, but still typically at least an order of magnitude smaller than one in realistic cases.   

Turning to disk impact, we compare the advective velocity $U_l(\tau)$ to $U_0$ at the semi-classical marginal onset time $\tau=\tau_\textrm{marg,sclass}$ (Eq.~\eqref{taumargsclass_disk})
\begin{equation}\label{eq:WIliquidtoU0}
\frac{U_l(\tau_\textrm{marg,sclass})}{U_0} = \frac{\sqrt{2}\,\delta}{(1-\delta)^{3/4}(1+\delta)^{1/2}w(\delta)^{1/2}} 
\frac{\left(\sigma g /\rho_l\right)^{1/4}}{U_0} 
\approx \sqrt{2\,\delta} \frac{U_\textrm{cap}}{U_0}  \qquad(\textrm{for}\,\,\delta\ll1)\,,
\end{equation}
where $U_\textrm{cap} \equiv (\sigma g /\rho_l)^{1/4}$ is the capillary velocity scale well-known from the minimum propagation speed of waves on a liquid interface, to which it is proportional. For the water-air interface we find that $U_\textrm{cap} \approx 0.164$ m/s, and therefore we may conclude that also in this case under typical impact conditions where one would expect instabilities to be of importance ($U_0 > 1.0$ m/s), the advective velocity is at least an order of magnitude smaller than $U_0$. A similar conclusion can be drawn for the wave speed. 

In conclusion, the advection and wave speeds within the emerging patterns are expected to be small: They are a factor $\delta$ smaller than the diverging gas speed and at marginal instability onset time they are at least an order of magnitude smaller than the approach speed $U_0$.

\section{Viscous and compressibility effects}
\label{sec:viscous}

For future comparison to experiments, it is important to see the significance of the results obtained in the previous Section in the light of other material properties. We will briefly discuss two of them in the present Section, namely the influence of gas viscosity and that of gas compressibility. Both of them will act as a limiting factor on the divergence of the gas flow speed, and the main objective will be to investigate which event takes place first, the instability onset or the curbing of the gas flow.  

\subsection{The influence of gas viscosity} 
\label{subsec:gasviscosity}

\begin{figure}
\centerline{\includegraphics[width=\textwidth]{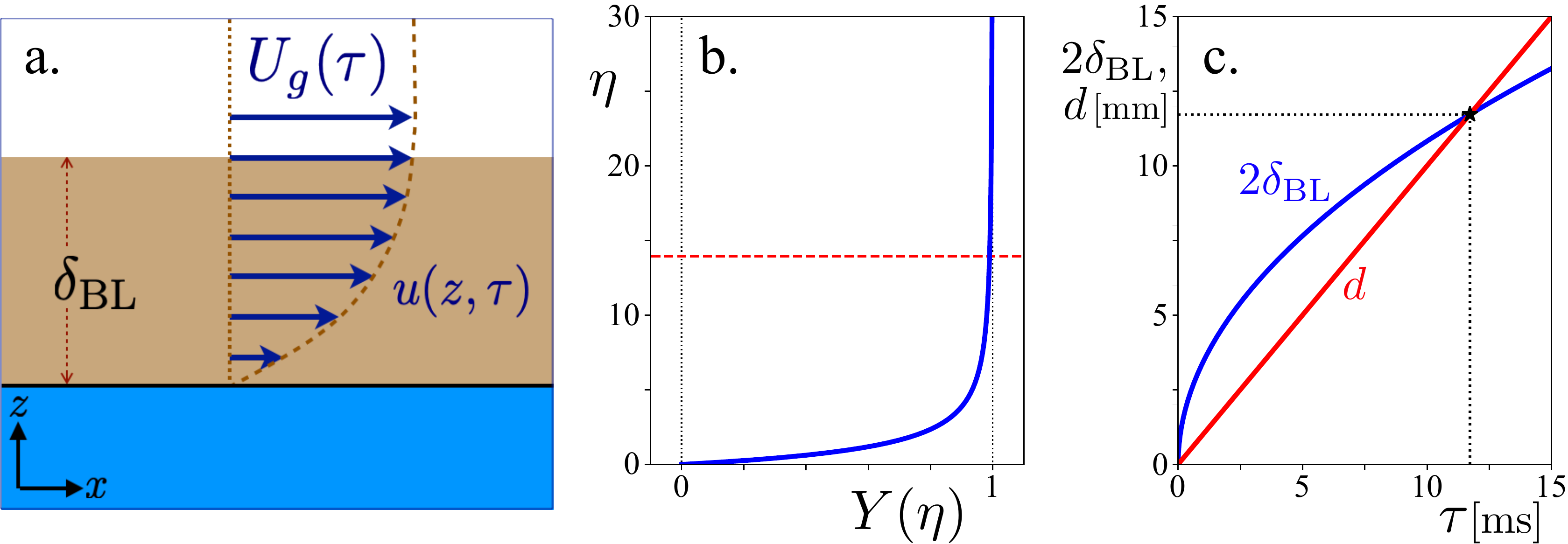}}% Images in 100% size
\caption{a. Schematic of the situation calculated in Appendix~\ref{appE}, where a boundary layer of thickness $\delta_\textrm{BL}$ is forming above a flat (liquid or solid) interface due to a gas flow of velocity $U_g(\tau)$ that diverges for $\tau \to 0$ in the horizontal ($x$) direction. The velocity in the boundary layer $u(z,\tau)$ is assumed to be parallel to the interface and only to depend on $z$ and $\tau$. b. Solution of the dimensionless boundary layer profile $Y(\eta) = u(z,\tau)/U_g(\tau)$ as a function of $\eta = z/\sqrt{\nu\tau}$. The horizontal red dashed line at $\eta_\textrm{BL}$ indicates the size of the boundary layer. c. Twice the boundary layer thickness $2\delta_\textrm{BL} = 2\sqrt{\nu\tau}$ together with the gap width $d = U_0\tau$, both as a function of $\tau$. The intersection of the two curves (black asterisk) indicates the time $\tau$ beyond which the system is dominated by gas viscosity. For these curves, $\nu = 15.1\cdot10^{-6}$ m$^2$/s corresponding to air at 20 $^\circ$C and atmospheric pressure, and $U_0 = 1.0$ m/s.      
}
\label{fig:appE}
\end{figure}

To characterize what happens at the (undisturbed and flat) gas-liquid interface due to the presence of gas viscosity, we imagine solving for a laminar flow that, far away from the interface, is identical to the gas flow we studied in the earlier Sections of this work,  $U_g(\tau) = q_0/(U_0\tau)$. We somewhat simplify the problem by assuming that the liquid remains stationary during the development of this boundary layer, which implies that the expected small motions within the liquid are neglected. This turns the interface into a flat and rigid boundary, and we aim at solving the laminar boundary layer flow that forms above this interface. Due to translational invariance we may assume that this flow is described by a flow field of the form $\vec{u} = u(z,\tau)\vec{e}_x$, with boundary conditions $u(0,\tau)=0$ and $u(\infty,\tau) = U_g(\tau)$, as depicted in Fig.~\ref{fig:appE}a. Here, $\nu$ represents the kinematic viscosity of the gas. 

If the gas velocity would be constant, the problem would correspond to the famous textbook problem of the impulsively started plate, in a moving reference frame in which the plate is at rest, however, where the analytic solution is an error function. In fact, for the above form of $U_g(\tau)$, we can find an analytic, self-similar solution for the time-dependent problem as well, when we start at some very far initial time $\tau_0$ that does not influence the system at the time $\tau$ of interest any longer:
\begin{equation}\label{eq:eqE_result}
\frac{u(z,\tau)}{U_g(\tau)} = \tfrac{1}{2}\sqrt{\pi}\,\left(\frac{z}{\sqrt{\nu\tau}}\right)\,\exp\!\left[\frac{z^2}{4\nu\tau}\right]\erfc\!\left[\frac{z}{2\sqrt{\nu\tau}}\right] \,,
\end{equation}
which expression is derived in Appendix~\ref{appE} and plotted in Fig.~\ref{fig:appE}b. Most important for the purpose of this Section however is that the boundary layer thickness $\delta_\textrm{BL}(\tau)$ is given by  
\begin{equation}\label{eq:eqE_BLwidth}
\delta_\textrm{BL}(\tau) = 13.93 \sqrt{\nu\tau}\,,
\end{equation} 
where the prefactor $13.93$ is determined from the point at which $u(z,\tau)=0.99U_g(\tau)$, or, equivalently, $Y(\eta)=0.99$, indicated by the horizontal dashed red line in Fig.~\ref{fig:appE}b. Note that approaching the impact moment (i.e., for $\tau\to0$), the boundary layer \emph{decreases} in size, which sounds counterintuitive at first sight, for those familiar with the behavior of viscous boundary layer development for steady asymptotic conditions where the boundary layer thickness is always an increasing function of time. It can however be understood intuitively by realizing that boundary layer formation needs to keep up with the divergence of the outer flow $U_g(\tau)$, where for equal subsequent steps in time, the steps in velocity diverge as well, which makes the boundary layer thickness shrink in time. 

However, even if the boundary layer thickness decreases in time, it does so as the square root of $\tau$, and as a consequence is at some point in time overtaken by the thickness of the gap between the liquid and the solid, which decreases linearly in time, as $d(\tau) = U_0\tau$. In Fig.~\ref{fig:appE}c, we compare twice the boundary layer thickness $\delta_\textrm{BL}(\tau)$ with the gap thickness $d(\tau)$ and see that there is the expected intersection point at 
\begin{equation}\label{eq:tauvisc}
\tau_\textrm{visc} = 776 \frac{\nu}{U_0^2}\,,
\end{equation} 
where for all $\tau<\tau_\textrm{visc}$ the viscous boundary layers at the two sides of the gap will touch and a viscous Poiseuille flow will start to form with a rapidly increasing pressure head and decreasing velocity that will prevent the type of instabilities discussed in the previous Section from occurring. On the other hand, if an instability develops at a time $\tau$ before $\tau_\textrm{visc}$, the fact of its growth is not likely to be hindered substantially by the action of gas viscosity. 

A short comment is in place since in the above calculation a no-slip boundary condition has also been assumed on the gas-liquid interface, whereas a no-shear boundary condition would be more appropriate. This is however justified as long as the dynamic viscosity in the liquid phase is much larger than that in the gas phase, which is satisfied for many gas-liquid combinations, e.g., air and water at atmospheric conditions.

\begin{figure}
\centerline{\includegraphics[width=\textwidth]{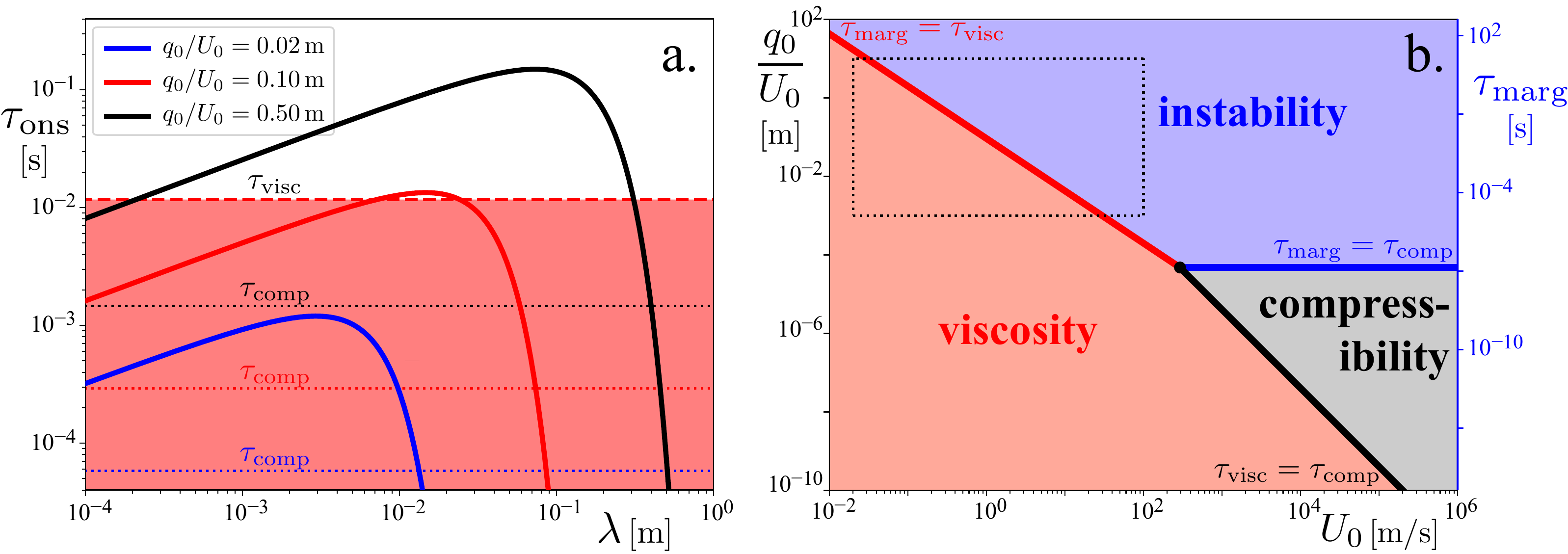}}
\caption{a. Doubly logarithmic plot of the onset time $\tau_\textrm{ons}$ versus wavelength $\lambda$ for the case of a wave impacting a vertical wall, identical to the one in Fig.~\ref{fig:WaveImpact}, i.e., for the same three different values of the gas flow rate parameter $q_0/U_0$ (solid lines) and $p=1.5$. The horizontal dashed black line represents the threshold time $\tau_\textrm{visc}$ beyond which the flow in the gap becomes dominated by viscosity (red shaded area), and the horizontal dotted lines indicate the threshold times $\tau_\textrm{comp}$ beyond which the flow is choked, for the three values of $q_0/U_0$, respectively. Here, we also fixed $U_0 = 1.0$ m/s.
b. Phase diagram, stating in what regions of the control parameter space of the wave impact problem, spanned by the length scale $q_0/U_0$ and velocity scale $U_0$, instabilities may occur, or viscosity or compressibility effects prevent the occurrence of an instability. The dotted black square indicates the region where experimental situations are likely to occur. See text for further explanation.  
Densities, interfacial tension, gas viscosity and gas speed of sound correspond to those of water and air at $20$ $^\circ$C and atmospheric pressure ($\rho_l = 998$ kg/m$^3$, $\rho_g = 1.20$ kg/m$^3$, $\sigma = 0.073$ N/m, $\nu = 1.51\cdot10^{-5}$ m$^2$/s, and $c_g = 343$ m/s.)       
}
\label{fig:phasediagram}
\end{figure}

\subsection{The influence of gas compressibility} 
\label{subsec:gascompressibility}

As the gas flow speed $U_g(\tau)$ is diverging, at some point in time it will get close to sonic values, i.e., it will become equal to the speed of sound in the gas, $c_g$. Taking the Mach number $\textit{Ma}$ of order unity as the threshold, $\textit{Ma} = U_g/c_g = 1$ we find
\begin{equation}\label{eq:taucomp}
\tau_\textrm{comp} = \frac{q_0}{U_0\,c_g}\,.
\end{equation}    
For all $\tau<\tau_\textrm{comp}$, the gas velocity in the gap becomes supersonic, and the flow will become choked which will cause the pressure in the air pocket (or centre of the disk) to rise, and simultaneously prevent an instability as from the previous Section to occur. As in the above discussed  case, if an instability arises for  $\tau>\tau_\textrm{comp}$ the subsequent choking of the flow may influence but not prevent the development of the instability.\\

Now let us estimate, for a moderate impact speed $U_0 = 1.0$ m/s, what the influence of gas viscosity and gas compressibility is on the stability diagram for the wave impact case (Subsection~\ref{subsec:waveimpact}), where we have used the properties of water and air at atmospheric pressure and a temperature of $20$ $^\circ$C. In Fig.~\ref{fig:phasediagram}a, we again plot the onset time as a function of the wavelength, but we now add the time $\tau_\textrm{visc}$ beyond which viscous effects prevent the development of a Kelvin-Helmholtz instability (dashed horizontal black line), together with the times $\tau_\textrm{comp}$ beyond which this happens due to choked flow (dotted horizontal lines). Since $\tau_\textrm{comp}$ depends on $q_0/U_0$ there are three of those. Clearly, $\tau_\textrm{comp}$ is smaller than $\tau_\textrm{visc}$ and the maximum $\tau_\textrm{marg}$ of the $\tau_\textrm{ons}(\lambda)$-curves for all values of $q_0/U_0$, such that we can conclude that gas compressibility is not an issue for the parameter settings of Fig.~\ref{fig:phasediagram}a. This is however not the case for gas viscosity: For the lowest value of $q_0/U_0$ ($= 0.02$ m) the entire instability onset curve lies within the red shaded area, and the occurrence of an instability is therefore out of the question. For the intermediate value ($q_0/U_0 = 0.10$ m), only the maximum itself lies outside the red shaded area, and it seems doubtful that an instability will occur. Only for the highest value,  $q_0/U_0 = 0.50$ m, a substantial part of the instability onset curve lies above the  $\tau_\textrm{visc}$-threshold and an instability is expected to be observed. 

Finally we explore in what regions of the direct control parameter space of the wave impact problem, spanned by the length scale $q_0/U_0$ and velocity scale $U_0$, Kelvin-Helmholtz instabilities may occur, and where viscosity or compressibility effects prevent the occurrence of such an instability. To that end, for each point in this parameter space, we look at the ordering of the three time scales $\tau_\textrm{marg}$, $\tau_\textrm{visc}$, and $\tau_\textrm{comp}$, provided by Eqs. \eqref{eq:taumarg}, \eqref{eq:tauvisc}, and \eqref{eq:taucomp}, respectively. E.g., if $\tau_\textrm{marg} > \tau_\textrm{visc},\tau_\textrm{comp}$, then instability is expected to occur. If however $\tau_\textrm{visc} > \tau_\textrm{marg},\tau_\textrm{comp}$, viscous effects prevent the development of an instability, etc. The resulting phase diagram is plotted in Fig.~\ref{fig:phasediagram}b. The lines are computed by time and again equating two of the three representative time scales, and all three intersect in a triple point, the coordinates of which can be computed as 
\begin{equation}\label{eq:triplepoint}
\left(U_0\,,\,\frac{q_0}{U_0}\right) = \left(h(p)\,n(\delta)\,\sqrt{776\,\frac{\rho_l\nu c_g^3}{\sigma}} \,\,,\,\, \frac{\sigma}{\rho_l \,(h(p)\,n(\delta)\,c_g)^2}\right)
\end{equation}    
The lines separate the regions where instabilities are expected, where they are impeded by viscosity, and where by gas compressibility. The most interesting area is indicated by the dotted black line in the form of a rectangle, which provides the bounds of situations that may practically occur. That is, for water and air, the competition of gas viscosity and Kelvin-Helmholtz instability dominate the scene, whereas gas compressibility is expected to play a very minor role, simply because it is restricted to the simultaneous incidence of very small $q_0/U_0$ and very large $U_0$, highly unlikely to be found in an actual experimental situation. However, examining Eq.~\eqref{eq:triplepoint}, one observes that for a (e.g. cryogenic) gas with a lower speed of sound the triple point would move towards or even into the area of interest.   

The general shape of the phase diagram~\ref{fig:phasediagram}b is understood intuitively when one realizes that $q_0/U_0$ is the main parameter regulating the instability onset time. That is, for large $q_0/U_0$ instabilities occur before viscosity and compressibility effects become relevant. For small $q_0/U_0$ viscosity and compressibility do become important before instability onset, and, clearly, for small values of the impact speed $U_0$ viscosity dominates, whereas for large impact speeds compressibility is the determining factor. From a more quantitative point of view, it is quite remarkable that gas viscosity is of very substantial importance, and even capable of stabilizing interfaces at the considerable length scale of $\sim 0.10$ m, whereas gas compressibility turns out to play a minor role. 

A final word of caution is appropriate because continuum theory and with that all of the above reasoning breaks down at the Knudsen limit, i.e., when the distance between the impacting solid and liquid becomes of the order of the mean free path $\ell$. For an ideal gas,  $\ell = k_BT/(\sqrt{2}p\Sigma)$ with $k_B$ Boltzmann's constant and $\Sigma$ the collisional cross-sectional area. For air at 20 $^\circ$C and atmospheric pressure, $\ell \approx 70$ nm, leading to Knudsen times $\tau_K = \ell/U_0$ which are significantly smaller than the other time scales and are situated in the lower left corner of Fig.~\ref{fig:phasediagram}b and are therefore irrelevant. However, the Knudsen limit may become an issue at lower ambient pressures.

\section{Comparison with experiments}
\label{sec:comparison}

\begin{figure}
\centerline{\includegraphics[width=0.65\textwidth]{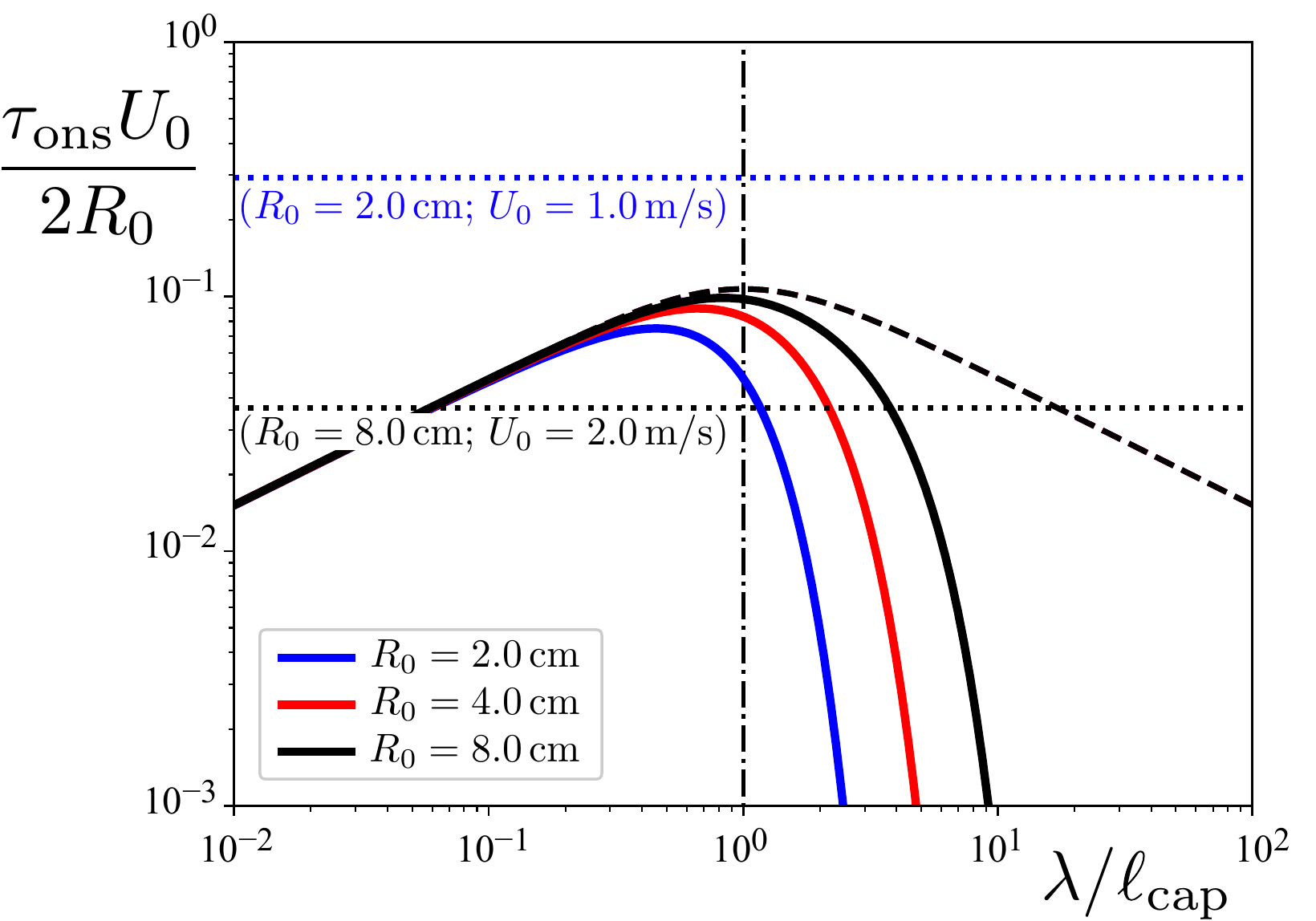}}
\caption{Doubly logarithmic plot of the dimensionless onset time $\tau_\textrm{ons}U_0/(2R_0)$ (with $U_0 = 1.0$ m/s) versus dimensionless wavelength $\lambda/\ell_\textrm{cap}$ (with $\ell_\textrm{cap} = 2\pi\sqrt{\sigma/[(1-\delta)\rho_l g]}$) for disk impact, using disk sizes in the range used in the experiment from \cite{jain2020} ($R_0 = 2.0, 4.0, 8.0$ cm, solid lines) and a moderately small value of the threshold $p=1.1$. As described in the text, the parameter $B$ is adjusted to provide instability at a lower, experimentally observed gas velocity. The black dashed line corresponds to the semi-classical result. The horizontal dotted lines give the times below which viscous effects in the gas layer will become dominant for $R_0 = 2.0$ cm, $U_0 = 1.0$ m/s (blue) and for $R_0 = 8.0$ cm, $U_0 = 2.0$ m/s (black). 
Densities, interfacial tension and gas viscosity correspond to the experimental conditions (water and air at $20$ $^\circ$C and atmospheric pressure): $\rho_l = 998$ kg/m$^3$, $\rho_g = 1.20$ kg/m$^3$, $\sigma = 0.073$ N/m, and $\nu = 1.51\cdot10^{-5}$ m$^2$/s.    
}

\label{fig:exponsetconditions}
\end{figure}

Unfortunately, to our knowledge, currently few experimental observations of the occurrence of instabilities in a diverging flow are available in the literature and most of them do not focus on this particular effect in a systematic way.

Nevertheless, a few interesting qualitative and quantitative results may be found that are at least consistent with the analysis presented here. The first are investigations in the impact of breaking waves generated in a flume on different scales, as discussed by \cite{lafeber2012a,lafeber2012b,bogaert2019}. From this work it is clear that instabilities become more prominent when the scale of the wave increases, which is consistent with the result of Subsection~\ref{subsec:waveimpact}, where it is predicted that both the marginal wavelength and the marginal onset time of the instability increase with the length scale $q_0/U_0$. 
%Also, unpublished results suggest that instabilities become more prominent at higher gas-to-liquid density ratios $\delta$, which is qualitatively consistent with the increase of the marginal onset time as $\delta^{3/4}$.
      
The second is the occurrence of a Kelvin-Helmholtz-type instability reported in our own experimental work \citep{jain2020} under the edge of a flat disk impacting on a water surface. There, for disk radii ranging from $R_0 = 1.5$ to $8.0$ cm, we reported the formation of an upward bulge with a disk-size independent marginal wavelength (that is, not directly scaling with the base flow in the gas), in the range predicted by classical Kelvin-Helmholtz stability analysis. Only for the largest disk sizes this bulge was evident in the surface deformation itself, but it was observable in the growth rate of the interface, i.e., also for the smaller disk sizes. In addition, it was observed that the instability did not occur at the marginal gas velocity predicted by the classical analysis ($U_{g,\textrm{marg}} = 6.58$ m/s), but at an earlier point in time, corresponding to $U_{g,\textrm{marg}} = 2.3$ m/s, which is a factor $2.9$ smaller.

We may account for the above factor by multiplying $B \sim U_g^2$ with the square of this factor (i.e. $2.9^2 = 8.4$, such that in the case where the semi-classical result is expected, the instability onset occurs at the earlier point in time where $U_{g,\textrm{marg}} = 2.3$ m/s. 

In Fig.~\ref{fig:exponsetconditions} we plot the stability diagram for three disk radii representative for the experiment, with the above correction in $B$ and 
%for a very small value of the threshold $p=1.01$
for the same threshold value $p=1.1$ that was used in Fig.~\ref{fig:DiskImpact}, and in addition provide the time threshold below which gas viscosity becomes dominant, for two different values of the disk impact speed. Since we non-dimensionalize the onset time $\tau_\textrm{ons}$ using the inertial time scale $2R_0/U_0$, all curves collapse with the semi-classical result for small wavelengths and start to deviate in the vicinity of the maximum. Since the deviations from the semi-classical result are minor, one may expect to find instabilities that emerge at the same dimensionless onset time of approximately $0.1$, which is consistent with the experimental onset value $\tau_\textrm{ons}U_0/2R_0 \approx 0.11$ reported in \cite{jain2020}.

The horizontal dotted lines provide the times below which viscous effects in the gas layer will become important, where the upper, blue one corresponds to $R_0 = 1.5$ cm and $U_0 = 1.0$ m/s and the lower, black one to $R_0 = 6.0$ cm and $U_0 = 2.0$ m/s. This suggests that results are at least somewhat affected by the action of viscosity, which, especially for the smallest disk sizes, may explain why in experiment the instability is not clearly observable as a deformation of the interface, but only from its growth rate. 

This is also consistent with the observation of \citet{mayer2018}, who for impacts of a rectangular flat plates with a maximal half width of $R_0 = 2.5$ cm and impact velocities up to $U_0=1.0$ m/s, upon visual inspection did not observe any upward deformation of the free surface at the rim: For the measured upper bounds the viscous onset time satisfies $\tau_\textrm{visc} \geq 12$ ms, whereas for the onset time we find $\tau_\textrm{ons} \leq 2$ ms. Because viscous effects become important at an earlier point in time (larger $\tau$), in the parameter space studied by \citet{mayer2018} no instability is expected to occur.
%FROM Mayer2018, page 1097:
%from rectangular plates of a half-width w_D 5?25 mm and total length l_D 70 mm,
%FROM Mayer2018, page 1097:
%At the start of the plate
%motion, far from the liquid surface, see figure 24(a), there is no significant deflection
%of the free surface. Although the l
ocation of initial contact in figure 24(b) cannot be
%accurately measured from the front view images owing to the front and back wall
%menisci, from the underside of the plate we can observe that no other points of contact
%are made except for at the plate edge.
%

Furthermore, Kelvin-Helmholtz-type instabilities had been observed in a larger scale study \citep{oh2009} where a large rectangular impactor (of $30$ cm width) had been impacted onto a water surface. In this case, multiple spatially periodic ripples localized under the impactor's edge were observed, but were however not further quantified. Also the study of \cite{ermanyuk2011} indicates the presence of concentric ring-shaped arrays of bubbles below an impacting flat disk with diameter of $18$ cm, but additional research is required to connect this observation to a Kelvin-Helmholtz-type instability.

%2) Talking about comparison with experiments, in figure 4 (upper row) of Ermanyuk \& Gavrilov (J. of Appl. Mech. \& Techn. Phys, Vol. 52, No. 6, pp. 889?895, 2011) there seem to be a (vague) indication that some preferable wavelength is indeed selected by the system: there are ring-shaped arrays of bubbles visible in image (b) located at more or less regular distance along the radial coordinate. However, it is unclear to which extent this pattern could be affected by acceleration (since it is a free-fall and not a constant-velocity experiment).

Finally, we have compared our theoretical predictions with the work of \citet{meerkerk2020b}, who developed a stereo planar laser-induced fluorescence technique \citep{meerkerk2020a} to measure the local wave shape in a flume, with which they were able to measure the local shape of the crest of a breaking wave in a wave flume as it approaches the wall upon which it impacts, and report measurements for $12$ repetitions of the same experimental conditions. Since they characterized both the time evolution of the wave shape and the entrapped air pocket extremely well, the necessary quantities can be obtained with precision, namely the radius of curvature at the crest ($R_c = 17$ mm), the impact velocity of the crest ($U_0 = 2.7$ m/s) and the volumetric flow rate in the neck ($q_0 \approx 0.40 \pm 0.05$ m$^2$/s). The last of these quantities has a relatively large error since it is not constant, but decreases as the impact approaches. Inserting $q_0$ and $U_0$ into the expressions~\eqref{eq:lambdamarg} and~\eqref{eq:taumarg} for the marginal wavelength and the marginal onset time, we obtain $\lambda_\textrm{marg} = 2.2$ cm and $\tau_\textrm{marg} = 25$ ms, where $\lambda_\textrm{marg}$ is in fair agreement with the size of the disturbances that were observed to develop in the crest region, which are however already visible in the first frame provided, at $\tau = 28$ ms, so even slightly earlier than the predicted onset time. Note that for the impact velocity $U_0 = 2.7$ m/s, the viscous onset time equals $\tau_\textrm{visc} = 1.6$ ms, so viscous effects are only expected to become important long after the instability has occurred. 

\section{Conclusion and discussion}
\label{sec:conclusion}

In this theoretical study we addressed the stability of a gas-liquid interface subject to a diverging flow speed in the gas layer such as will typically occur during the impact of a liquid onto a solid, with prime examples being the impact of a %(almost) 
flat plate onto a liquid surface, or the impact of a breaking wave onto a wall. In a simplified geometrical setting, leading to a gas velocity $U_g$ that increases inversely proportional to the amount of time $\tau$ remaining until impact, we formulated a linear stability analysis of the basic potential flow, including gravity and surface tension, but neglecting viscous effects in both phases. A modal decomposition subsequently led to an amplitude equation, which we showed to posses solutions with constant amplitude for large $\tau$ that however all diverge for sufficiently small $\tau$. We focussed on the growth rate and the onset time of divergence to assess to what extent solutions would lead to physically observable instabilities, given the fact that growth rate varies wildly between solutions and the fact that the amount of time allowing instabilities to grow is limited by the impact.  

The most significant general conclusion that can be drawn based on this work is that large wavelengths are stabilized by the diverging character of the flow. This stabilization is not so much caused by limiting the amount of time that the instabilities have to grow, since what we have called the semi-classical onset time may in fact even diverge for infinitely large wavelength (see, e.g., Fig.~\ref{fig:WaveImpact}), but more importantly by a strong suppression of the growth rate at these large wavelengths. The key parameter identified here is the dimensionless quantity $B = w(\delta)(q_0/U_0)^2k^2$, which depends on the density ratio $\delta$, the gas flow parameter $q_0/U_0$ and the wave number $k$, where the suppression happens as soon as $B$ is smaller than $1$, i.e., for sufficiently large wavelengths.   

To illustrate the above, we studied two physical examples, namely (i) the parallel impact of a wave onto a vertical wall and (ii) the impact of a horizontal plate onto a liquid surface. In the first case, where the influence of gravity is negligible due to the vertical nature of the problem, we find that the diverging nature of the classical (i.e., constant gas speed) Kelvin-Helmholtz stability diagram for long wavelengths is completely changed by taking into account the time-dependence: Instead of being divergent for long wavelengths, the diagram now exhibits a clear maximum in $\tau$, corresponding to the first wavelength to become unstable, i.e.,the marginal wavelength that is expected to show up in an experiment. We determined approximate expressions for this marginal wavelength and the corresponding marginal instability onset time in terms of the control parameters.  

In the second case, we find that for large values of the disk radius $R_0$ (where $q_0/U_0=R_0/2$), the stability diagram resembles that of classical situation: The marginal wavelength is solely determined by gravity, liquid and gas properties (Eq.~\eqref{lambdamargsclass_disk}), and independent of the gas flow rate, which only shows up in the marginal instability onset time. There is a sharp cut-off in the diagram for larger wavelength, that for smaller disk radii starts to interfere with the classical maximum: For small values of $R_0$ the instability is strongly suppressed and the location of the maximum shifts to smaller wavelengths (Fig~\ref{fig:DiskImpact}).           

Finally, we discussed the relevance of gas viscosity and gas compressibility. For the former, the well-known impulsively started plate plate problem was modified by incorporating the diverging gas flow speed, leading to an analytical prediction of boundary layer thickness converging as $\sim \sqrt{\nu\tau}$. We found that whereas gas compressibility has a mostly insignificant influence on instability onset, the reverse is true for gas viscosity: For water and air, gas viscosity is able to prevent the occurrence of an instability on length scales as large as $q_0/U_0 \sim 0.10$ m. Note that this does not imply that gas compressibility is unimportant at all, but that its influence is expected to occur after onset of instability, when the gas flow speed becomes of the order of the sound speed.

Two obvious simplifications have remained largely untouched in this work. The first is the particular form chosen for the divergence of the gas flow speed (i.e., $U_g \sim 1/\tau$), which apart from being physically plausible, also has the advantage of allowing for quite a number of analytical approximations. Clearly, the general picture will remain the same if $U_g(\tau)$ diverges in another manner, but the analysis needs to shift to a %somewhat 
more numerical approach rather than the more analytic route taken here. 

The second, and possibly more significant one, is the fact that for our analysis we have assumed the gas domain to be unbounded, whereas at some point impactor and target should get close enough to one another to be of influence. From a previous analysis in the context of classical Kelvin-Helmholtz stability theory \citep{jain2020}, we concluded for the disk case that there is no influence of the finite gap thickness as long as the gap width $d(\tau_\textrm{ons}) = U_0\tau_\textrm{ons}$ at the onset time is larger than the reciprocal value of the marginal wavenumber, i.e., $d(\tau_\textrm{ons}) > 1/k_\textrm{marg} = \lambda_\textrm{marg}/2\pi$. In general, the finite gap thickness is found to have a stabilizing effect on smaller wavelengths, but a more quantitative assessment of this effect requires further study.    

Experimental observations of instabilities occurring during slamming impacts are scarce, and usually of a rather qualitative and descriptive nature. Nevertheless, what is available is at least consistent with the analysis presented here and hopefully the current work will inspire more quantitative experimental research in the near future.

\section*{Acknowledgements}
The author acknowledges the many helpful and stimulating discussions with Hannes Bogaert, Laurent Brosset, Rodrigo Ezeta Aparicio, and Utkarsh Jain. The author acknowledges financial support from SLING (project number P14-10.1), which is partly financed by the Netherlands Organisation for Scientific Research (NWO). 

\section*{Declaration of Interests} 
The author reports no conflict of interest.

\appendix

%%% APPENDIX A %%%

\section{Amplitude equation for three-dimensional perturbations}
\label{appA}

In Section~\ref{sec:problem} we derived the amplitude equation for small perturbations of the interface in the $x$-direction. For completeness, we provide the expressions that are obtained when the interface is disturbed in any arbitrary direction in the horizontal plane %$(x,y)$-plane 
below. If $(x_1,x_2)$ denote the coordinates in the horizontal plane, $\vec{k} = (k_1,k_2)$ is the wave vector corresponding to the disturbance and $\vec{U}_g(\tau) = U_g(\tau)\vec{e}_1$ is the gas velocity, where the $x_1$-direction (with corresponding unit vector $\vec{e}_1$) has conveniently been chosen parallel to the gas velocity, then the main difference is that whereas the gravity, and surface tension related terms all involve the full wavenumber $k=(k_1^2 + k_2^2)^{1/2}$, whereas the terms that include the gas (and liquid) velocity involve only the $x_1$-component of the wave number, i.e., $k_1$. 

More specifically, in Eqs.~\eqref{eq:Laplace}-\eqref{eq:dynBC}, the only change is an additional term $\sigma(\partial^2\eta/\partial x_2^2)$ in the dynamic boundary condition (Eq.~\eqref{eq:dynBC}). The expressions for the disturbance in the surface and potentials thus become $\eta(x_1,x_2,t)=\varepsilon(t)\exp[ik_1x_1+ik_2x_2]$, $\phi_l^{(1)} = C_l(t)\exp[kx_3+ik_1x_1+ik_2x_2]$ and $\phi_g^{(1)} = C_g(t)\exp[-kx_3+ik_1x_1+ik_2x_2]$, where $x_3$ denotes the vertical coordinate. This then leads to 
\begin{eqnarray}
(\dot{\varepsilon} + ik_1\bar{U}_l\varepsilon) &=& kC_l\,,\label{eqA:kbc1}\\
(\dot{\varepsilon} + ik_1\bar{U}_g\varepsilon) &=& -kC_g\,,\label{eqA:kbc2}\\
\rho_l(\dot{C}_l + ik_1\bar{U}_lC_l +g\varepsilon) &=& \rho_g(\dot{C}_g + ik_1\bar{U}_gC_g +g\varepsilon) - \sigma k^2 \varepsilon\,. \label{eqA:dbc}
\end{eqnarray}  
Multiplying the last equation with $k$ and substituting $kC_l$ and $kC_g$ from the first two one arrives at an equation similar to Eq.~\eqref{eq:governing1}, which then with a similar transformation as in Eq.~\eqref{eq:modified}, but now involving only the $x_1$-component of the wave number $k_1$, leads to the amplitude equation for three-dimensional perturbations
\begin{equation}\label{eqC:governing2}
\frac{d^2\bar{\varepsilon}}{dt^2} + \left[\frac{1-\delta}{1+\delta}gk + \frac{\sigma}{\rho_l(1+\delta)}k^3 - \frac{\delta}{(1+\delta)^2}(U_g(\tau))^2 k_1^2\right]\bar{\varepsilon} = 0\,,
\end{equation}
where the difference with Eq.~\eqref{eq:governing2} is the presence of the full wave number $k$ in the gravity and surface tension terms and only the $x_1$-component $k_1$ in the gas-velocity dependent last term.

%%% APPENDIX C %%%

\section{Analytical solution of the amplitude equation}
\label{appC}

For completeness we write down the analytic solution to the amplitude equation~\eqref{eq:nondimampleq} discussed in Section~\ref{sec:amplitude}, by noting that the transformation $\tilde{y} = \sqrt{\tilde{\tau}}f(\tilde{\tau})$ maps Eq.~\eqref{eq:nondimampleq} onto the Bessel equation 
\begin{equation}\label{eqC:transf}
\tilde{\tau}^2\ddot{f} + \tilde{\tau}\dot{f} + (\tilde{\tau}^2-(B+\tfrac{1}{4}))f = 0\,,
\end{equation} 
which leads to the following general solution
\begin{equation}\label{eqC:generalsol}
\tilde{y} = C_J\sqrt{\tilde{\tau}}\,J_{\tfrac{1}{2}\sqrt{4B+1}}(\tilde{\tau})  +  C_Y\sqrt{\tilde{\tau}}\,Y_{\tfrac{1}{2}\sqrt{4B+1}}(\tilde{\tau}) \,,
\end{equation}
with $C_J$ and $C_Y$ integration constants and $J_s(\tilde{\tau})$ and $Y_s(\tilde{\tau})$ the Bessel functions of order $s$ of the first and second kind respectively. Other than that the above expression confirms that the solution is oscillatory for large $(\tilde{\tau})$ and diverges for small $(\tilde{\tau})$ it is of little practical value concerning the purposes of this article.

Also the complex amplitude growth rate $\tilde{\gamma}$ defined in Appendix~\ref{appB} may be expressed in terms Bessel functions, using recursive relations for the derivatives of the Bessel functions. For $F=J,Y$, we have, using the relation $dF_s/dz = -s/zF_s + F_{s-1}$
\begin{equation}\label{eqC:transf1}
\frac{d}{dz}\left(\sqrt{z}F_s(z)\right) = \frac{1}{2\sqrt{z}}F_s(z) + \sqrt{z}\frac{dF_s(z)}{dz} = \frac{1-2s}{2z} \sqrt{z}F_s(z) +  \sqrt{z}F_{s-1}(z)\,,
\end{equation} 
with which when applied to Eq.~\eqref{eqC:generalsol} we can use to express $\tilde{\gamma}$ in terms of Bessel functions
\begin{equation}\label{eqC:transf2}
\tilde{\gamma}(\tilde{\tau}) = -\frac{1}{\tilde{y}}\frac{d\tilde{y}}{d\tilde{\tau}} = \frac{\sqrt{4B+1}-1}{2\tilde{\tau}} + \frac{C_J\sqrt{\tilde{\tau}}\,J_{\tfrac{1}{2}\sqrt{4B+1}-1}(\tilde{\tau})  +  C_Y\sqrt{\tilde{\tau}}\,Y_{\tfrac{1}{2}\sqrt{4B+1}-1}(\tilde{\tau})}{C_J\sqrt{\tilde{\tau}}\,J_{\tfrac{1}{2}\sqrt{4B+1}}(\tilde{\tau})  +  C_Y\sqrt{\tilde{\tau}}\,Y_{\tfrac{1}{2}\sqrt{4B+1}}(\tilde{\tau})}\,.
\end{equation} 
The first term on the right hand side represents the divergence of $\tilde{\gamma}$ for $(\tilde{\tau}) \to 0$, but, again, otherwise the above expression has little practical value for this work.

%%% APPENDIX B %%%

\section{Evolution equation for the absolute amplitude growth rate $\tilde{\Gamma}$}
\label{appB}

The complex-valued, second-order amplitude equation~\eqref{eq:nondimampleq} discussed in Section~\ref{sec:amplitude} can be mapped onto a system of two coupled, real-valued, first-order equations for the two main quantities of interest in this stability analysis, namely the magnitude of the amplitude $|\tilde{y}|$ and the absolute amplitude growth rate $\tilde{\Gamma} \equiv (-1/|\tilde{y}|)d|\tilde{y}|/d\tilde{\tau}$.

To derive equations for the time evolution of these quantities, we start with defining the complex amplitude growth rate $\tilde{\gamma}$ as 
\begin{equation}\label{eqB:defgamma}
\tilde{\gamma} \equiv -\frac{1}{\tilde{y}}\frac{d\tilde{y}}{d\tilde{\tau}}\,.
\end{equation}
Taking the time derivative of $\tilde{\gamma}$ we arrive at
\begin{equation}\label{eqB:gammaequation}
\frac{d\tilde{\gamma}}{d\tilde{\tau}} = \left(\frac{1}{\tilde{y}}\frac{d\tilde{y}}{d\tilde{\tau}}\right)^2  -\frac{1}{\tilde{y}}\frac{d^2\tilde{y}}{d\tilde{\tau}^2} =  \tilde{\gamma}^2 + 1 - \frac{B}{\tilde{\tau}^2}\,,
\end{equation}
where in the second step we have used the evolution equation~\eqref{eq:nondimampleq} for $\tilde{y}$. Now we rewrite $\tilde{\Gamma}$ as
\begin{equation}\label{eqB:rewrGamma}
\tilde{\Gamma} = -\frac{1}{|\tilde{y}|}\frac{d|\tilde{y}|}{d\tilde{\tau}} = -\tfrac{1}{2}\frac{1}{|\tilde{y}|^2}\frac{d|\tilde{y}|^2}{d\tilde{\tau}} = -\tfrac{1}{2}\frac{1}{\tilde{y}\tilde{y}^*}\frac{d\tilde{y}\tilde{y}^*}{d\tilde{\tau}} \,,
\end{equation}
where $\tilde{y}^*$ denotes the complex conjugate of $\tilde{y}$. Taking the time derivative of this equation we obtain
\begin{equation}\label{eqB:eqGamma1}
\frac{d\tilde{\Gamma}}{d\tilde{\tau}} = -\tfrac{1}{2}\left[\frac{1}{\tilde{y}\tilde{y}^*}\frac{d^2\tilde{y}\tilde{y}^*}{d\tilde{\tau}^2} - \frac{1}{(\tilde{y}\tilde{y}^*)^2}\left(\frac{d\tilde{y}\tilde{y}^*}{d\tilde{\tau}}\right)^2\right] = -\tfrac{1}{2}\left[\frac{1}{\tilde{y}\tilde{y}^*}\frac{d^2\tilde{y}\tilde{y}^*}{d\tilde{\tau}^2} -4\tilde{\Gamma}^2\right] \,.
\end{equation}
Now the first term can be rewritten as
\begin{eqnarray}\label{eqB:eqGamma2}
\frac{1}{\tilde{y}\tilde{y}^*}\frac{d^2\tilde{y}\tilde{y}^*}{d\tilde{\tau}^2} &=& \frac{1}{\tilde{y}}\frac{d^2\tilde{y}}{d\tilde{\tau}^2} + \frac{1}{\tilde{y}^*}\frac{d^2\tilde{y}^*}{d\tilde{\tau}^2} + 2 \left(\frac{1}{\tilde{y}}\frac{d\tilde{y}}{d\tilde{\tau}}\right) \left(\frac{1}{\tilde{y}^*}\frac{d\tilde{y}^*}{d\tilde{\tau}}\right)\nonumber\\
&=& -2\left(1-\frac{B}{\tilde{\tau}^2}\right) + 2\tilde{\gamma}\tilde{\gamma}^*\,,
\end{eqnarray}
where we have again used ~\eqref{eq:ampleq} and its complex conjugate. One may now be tempted to equate the last term with $2\tilde{\Gamma}^2$, but in spite of its name, $\tilde{\Gamma}$ is \emph{not} the magnitude of $\tilde{\gamma}$. In fact:
\begin{equation}\label{eqB:GammaisRegamma}
\tilde{\Gamma} = -\tfrac{1}{2}\frac{1}{\tilde{y}\tilde{y}^*}\frac{d\tilde{y}\tilde{y}^*}{d\tilde{\tau}} = -\tfrac{1}{2}\left[\frac{1}{\tilde{y}}\frac{d\tilde{y}}{d\tilde{\tau}} + \frac{1}{\tilde{y}^*}\frac{d\tilde{y}^*}{d\tilde{\tau}}\right] =  \tfrac{1}{2}\left[\tilde{\gamma} + \tilde{\gamma}^*\right] = \mathcal{R}\left[\tilde{\gamma}\right]\,,
\end{equation}
where $\mathcal{R}[z]$ denotes the real part of the complex quantity $z$. Defining the imaginary part of $\tilde{\gamma}$ as $\mathcal{I}[\tilde{\gamma}] \equiv -\tilde{Z}$, we may collect Eqs.~\eqref{eqB:eqGamma1} to~\eqref{eqB:GammaisRegamma} into an evolution equation for $\tilde{\Gamma}$
\begin{equation}\label{eqB:evoleqGamma}
\frac{d\tilde{\Gamma}}{d\tilde{\tau}} = \tilde{\Gamma}^2 - \tilde{Z}^2 + 1 - \frac{B}{\tilde{\tau}^2}\,.
\end{equation}
However, this equation still contains the unknown quantity $\tilde{Z}$, for which we also need to obtain an evolution equation. By subtracting the complex conjugate from Eq.~\eqref{eqB:gammaequation} from itself, we find that the term $1-B/\tilde{\tau}^2$ cancels out
\begin{equation}\label{eqB:evoleqZhlp}
\frac{d(\tilde{\gamma}-\tilde{\gamma}^*)}{d\tilde{\tau}} = \tilde{\gamma}^2 - (\tilde{\gamma}^*)^2 = (\tilde{\gamma}+\tilde{\gamma}^*)(\tilde{\gamma}-\tilde{\gamma}^*),.
\end{equation}
Now, since $\tilde{\gamma}+\tilde{\gamma}^* = 2\mathcal{R}[\tilde{\gamma}] = 2\tilde{\Gamma}$ and $\tilde{\gamma}-\tilde{\gamma}^* = 2i\mathcal{I}[\tilde{\gamma}] = -2i\tilde{Z}$, the above equation provides us with the sought-for evolution equation for $\tilde{Z}$
\begin{equation}\label{eqB:evoleqZ}
\frac{d\tilde{Z}}{d\tilde{\tau}} = 2\tilde{\Gamma}\tilde{Z}\,,
\end{equation}
which completes the set of equations. (Note that, knowing that $\tilde{\Gamma}=\mathcal{R}[\tilde{\gamma}]$, the derivation in Eqs.~\eqref{eqB:eqGamma1} to~\eqref{eqB:evoleqGamma} is equivalent to adding Eq.~\eqref{eqB:gammaequation} and its complex conjugate.) In order to solve the equations we need initial conditions for $\tilde{Z}$ and $\tilde{\Gamma}$ at some initial time $\tilde{\tau}_0 (\gg \sqrt{B})$. In order to obtain those, we turn to the approximate solution $\tilde{y}(\tilde{\tau}) \approx \exp[i\psi + i\tilde{\tau}]$ (Eq.~\eqref{eq:earlylimsol}) and determine $\tilde{\gamma}$ by taking the derivative  
\begin{equation}\label{eqB:gammaBC}
\tilde{\gamma} = -\frac{1}{\tilde{y}}\frac{d\tilde{y}}{d\tilde{\tau}} \approx - \exp[-i\psi - i\tilde{\tau}]\, i \exp[i\psi + i\tilde{\tau}] = -i\,,
\end{equation}
we find that $\tilde{\gamma}$ approximates the constant value $-i$ in that regime, from which we determine that $\tilde{\Gamma}(\tilde{\tau}_0) = 0$ and $\tilde{Z}(\tilde{\tau}_0) = 1$ are the required initial conditions. 

Subsequently, one may compare the evolution equation~\eqref{eqB:evoleqZ} for $\tilde{Z}$ and the definition of $\tilde{\Gamma}$:
\begin{equation}\label{eqB:Zandy1}
\tilde{\Gamma} = \tfrac{1}{2} \frac{1}{\tilde{Z}}\frac{d\tilde{Z}}{d\tilde{\tau}}\qquad\textrm{and}\qquad\tilde{\Gamma} = - \frac{1}{|\tilde{y}|}\frac{d|\tilde{y}|}{d\tilde{\tau}}\qquad\,.
\end{equation} 
Clearly, since the left hand side of both equations are equal, so are the right hand sides, or
\begin{equation}\label{eqB:Zandy2}
\log\left(\tilde{Z}^{-1/2}\right) = \log|\tilde{y}| + C\,,
\end{equation} 
where C is an integration constant. Since $\tilde{Z}(\tilde{\tau}_0) = |\tilde{y}|(\tilde{\tau}_0) = 1$, we infer that C = 0, and therefore identify $|\tilde{y}| = 1/\sqrt{\tilde{Z}}$. 

In conclusion, we have mapped our second order time-evolution equation \eqref{eq:nondimampleq} for the complex quantity $\tilde{y}$ in two first order equations for the (real and non-negative) quantities of interest namely the absolute amplitude growth rate $\tilde{\Gamma}$ and the magnitude of the amplitude $|\tilde{y}|$:
\begin{eqnarray}\label{eqB:finalequations}
\frac{d\tilde{\Gamma}}{d\tilde{\tau}} &=& \tilde{\Gamma}^2 - \tilde{Z}^2 + 1 - \frac{B}{\tilde{\tau}^2}\,,\\
\frac{d\tilde{Z}}{d\tilde{\tau}} &=& 2\tilde{\Gamma}\tilde{Z}\qquad\qquad\qquad\qquad\textrm{with:}\quad \left|\tilde{y}\right| = \frac{1}{\sqrt{\tilde{Z}}}  \,,
\end{eqnarray} 
complemented with the boundary conditions:
\begin{equation}\label{eqB:BCs}
\tilde{\Gamma}(\tilde{\tau}_0) = 0\qquad\textrm{and}\qquad \tilde{Z}(\tilde{\tau}_0) = 1\,,
\end{equation}
at some initial time $\tilde{\tau}_0 \gg \sqrt{B}$. In fact, all numerical solutions shown in this article for $\tilde{\Gamma}$ and $|\tilde{y}|$ are obtained by numerically solving the above equations, rather than numerically solving the original equation~\eqref{eq:nondimampleq} or using the analytical expressions in terms of Bessel functions from Appendix~\ref{appC}, as the former lead to the most accurate results.

Now, finally, one may also understand the oscillations that were obtained in the early time regime (cf. Fig.~\ref{fig:absyGammavsB}b). When we define $\tilde{\zeta} \equiv 1-\tilde{Z}$ we note that both  $\tilde{\zeta}$ and  $\tilde{\Gamma}$ are small close to $\tilde{\tau}_0$ such that we may linearize Eqs.~\eqref{eqB:finalequations} leading to
\begin{eqnarray}\label{eqB:linfinalequations}
\frac{d\tilde{\Gamma}}{d\tilde{\tau}} &\approx& 2\tilde{\zeta} - \frac{B}{\tilde{\tau}^2}\,,\\
\frac{d\tilde{\zeta}}{d\tilde{\tau}} &\approx& -2\tilde{\Gamma}  \,,
\end{eqnarray} 
which by taking the derivative of the first may be written as
\begin{equation}\label{eqB:linfinaleqs2}
\frac{d^2\tilde{\Gamma}}{d\tilde{\tau}^2} + 4 \tilde{\Gamma} \approx 2\frac{B}{\tilde{\tau}^3} \approx 0\,.
\end{equation}
The solution of this equation provides oscillations with (non-dimensional) frequency 2, i.e., a period-doubled modulation of $|\tilde{y}|$. Note that $|\tilde{y}| \approx 1 + \tfrac{1}{2}\tilde{\zeta}$ in this limit. 

%%% APPENDIX F %%%

\section{Numerical growth factor}
\label{appF}

Clearly, relevant solutions of the amplitude equation~\eqref{eq:ampleq} are diverging to $+\infty$ in the limit $\tau\to 0$. This seems to imply that within the context of the linear stability analysis all instabilities would grow indefinitely and as such make any quantitative prediction about to what extent a disturbance of a certain wave number may grow impossible.

There exists however a quite natural maximal value $|y|_\textrm{max}$ beyond which the amplitude can never grow, namely when it has become equal to the gap width that separates the undisturbed gas-liquid interface from the solid and an impact with the the maxima of the disturbance would take place. Written symbolically, $|y|_\textrm{max}$ is the solution of
\begin{equation}\label{eqF:ymax}
|y(\tau)| = U_0\,\tau\,. 
\end{equation}
Non-dimensionalizing using that $\tau = \tilde{\tau}/\sqrt{A}$ and $|y| = |\tilde{y}| y_0$, with $y_0$ the size of the initial disturbance, we can rewrite the above expression as
\begin{equation}\label{eqF:nondimymax}
\tilde{\tau} = \frac{\sqrt{A}\,y_0}{U_0}\,|\tilde{y}(\tilde{\tau})|\,,
\end{equation}
which expression we may solve using the numerical solution of the non-dimensional amplitude equation~\eqref{eq:nondimampleq} to obtain the maximum growth factor $|\tilde{y}|_\textrm{max}$, which describes by what multiplicative factor the initial disturbance has grown.

To express the answer as a function of the parameter $B$ ($= \beta_2 k^2$, with $k$ the wave number), we substitute $k = \sqrt{B/\beta_2}$ into $A(k)$. For wave impact ($A(k) = \alpha_3k^3$) this leads to
\begin{equation}\label{eqF:ymaxwaveimpact}
\tilde{\tau} = \sqrt{\frac{\sigma\rho_l^{1/2}}{\rho_g^{3/2}}}\sqrt{\frac{U_0}{q_0}}\frac{y_0}{q_0}B^{3/4} |\tilde{y}(\tilde{\tau})| \equiv \chi B^{3/4} |\tilde{y}(\tilde{\tau})|\,,
\end{equation}
whereas for disk impact ($A(k) = \alpha_1k + \alpha_3k^3$) the expression is slightly more involved
\begin{equation}\label{eqF:ymaxdiskimpact}
\tilde{\tau} = \sqrt{\frac{\sigma\rho_l^{1/2}}{\rho_g^{3/2}}}\sqrt{\frac{U_0}{q_0}}\frac{y_0}{q_0}B^{3/4} \sqrt{1 + \frac{\rho_gg}{\sigma}\left(\frac{q_0}{U_0}\right)^2B^{-2}} \,\,|\tilde{y}(\tilde{\tau})| \equiv \chi B^{3/4}\sqrt{1 + \Xi B^{-2}}\, |\tilde{y}(\tilde{\tau})|\,.
\end{equation}
Here we have used the expressions~\eqref{eq:Adef} and~\eqref{eq:Bdef} to express $\alpha_1$, $\alpha_3$, and $\beta_2$ in fluid properties and control parameters, and defined the dimensionless parameters $\chi$, which is proportional to the initial amplitude $y_0$, and $\Xi$. 

Using the properties of water and air at $20$ $^\circ$C and atmospheric pressure ($\rho_l = 998$ kg/m$^3$, $\rho_g = 1.20$ kg/m$^3$, and $\sigma = 0.073$ N/m), the largest value $q_0/U_0 = 0.50$ m plotted in Fig.~\ref{fig:WaveImpact}, a typical value of the impact speed $U_0 = 6.0$ m/s, and setting the initial amplitude to the (rather arbitrary but small) value of $y_0 = 25$ nm, we obtain for wave impact that $\chi = 0.622 y_0 = 1.56\cdot10^{-8}$. Numerically solving Eq.~\eqref{eqF:ymaxwaveimpact} then leads to the result plotted in Figs.~\ref{fig:TauonsVsB}b and \ref{fig:WaveImpact}.

For disk impact we use instead the middle value $q_0/U_0 = R_0/2 = 0.05$ m, an impact speed $U_0 = 2.0$ m/s, and the same initial amplitude $y_0 = 25\cdot10^{-9}$ m to compute $\chi = 59.0 y_0 = 1.48\cdot10^{-6}$ and $\Xi = 1.62$. The results plotted in Figs.~\ref{fig:TauonsVsB}b and \ref{fig:DiskImpact} correspond to the numerical solution of Eq.~\eqref{eqF:ymaxdiskimpact} with these parameter values.

It is possible to obtain an approximate analytic expression for the equation~\eqref{eqF:nondimymax} using the approximate asymptotic solution~\eqref{eq:approxsol} of the amplitude equation valid for $\tilde{\tau}<\sqrt{B}$. Solving for $|\tilde{y}|_\textrm{max}$ this leads to
\begin{equation}\label{eqF:ymaxwaveimpact}
|\tilde{y}|_\textrm{max,approx} = \left(\chi B^{1/4}\right)^L\quad\textrm{with}\,\, L = -\frac{\sqrt{4B+1}-1}{\sqrt{4B+1}+1}
\end{equation}
for wave impact, and to
\begin{equation}\label{eqF:ymaxdiskimpact}
|\tilde{y}|_\textrm{max,approx} = \left(\chi B^{1/4} \sqrt{1 + \Xi B^{-2}}\,\right)^L
\end{equation}
for disk impact. These expressions predict the numerical growth rate accurately for small $B$ ($\lesssim 5$) but overpredict for large $B$, which is connected to the fact that there the onset happens close to the dimensionless crossover time $\tilde{\tau} = \sqrt{B}$ where the numerical solution deviates strongly from the approximate one (cf. Fig.~\ref{fig:absyGammavsB}a). Finally, note that the exponent $L$ asymptotically tends to $-1$ for large $B$, leading to very large maximum growth factors $|\tilde{y}|_\textrm{max}$, and to $B \approx 0$ for small $B$, leading to $|\tilde{y}|_\textrm{max} \approx 1$.   

%%% APPENDIX D %%%

\section{Marginal wavelength and onset time}
\label{appD}

The marginal wavelength in the case of wave impact (Subsection~\ref{subsec:waveimpact}) is found straightforwardly by solving $d\tau_\textrm{ons,w}/dk = 0$ where $\tau_\textrm{ons,w}(k)$ is provided by Eq.~\eqref{tauons_wave}. The algebraically somewhat involved details of this calculation are presented here for completeness, including the definition of the functions $g(p)$, $h(p)$, $m(\delta)$, and $n(\delta$) of threshold $p$ and density ratio $\delta$ used in the main text. 

Computing the derivative and equating it to zero directly leads to the condition
\begin{equation}\label{eqD:cond1}
\log(p) = \frac{(\sqrt{4X+1}-1)^2\sqrt{4X+1}}{16X}\,, 
\end{equation}
where we have defined $X \equiv \beta_2k^2 (= B)$. From this expression we immediately see that we may express $X$ in terms of $p$, i.e., we may define a function $g(p)$ such that
\begin{equation}\label{eqD:g_p}
\sqrt{X} = g(p) \qquad\Rightarrow\qquad k_\textrm{marg} = \frac{g(p)}{\sqrt{\beta_2}} \,.
\end{equation}
To solve Eq.~\eqref{eqD:cond1} we transform $w \equiv \sqrt{4X+1}$, from which we find $X = (w^2-1)/4$ and write $s = \log(p)$. Herewith Eq.~\eqref{eqD:cond1} becomes
\begin{equation}\label{eqD:cond2}
s = \frac{(w-1)^2w}{4(w^2-1)} =  \frac{(w-1)w}{4(w+1)} \,,
\end{equation}
which leads to the quadratic equation $w^2 -(1+4s)w - 4s = 0$ which is readily solved for $w$ as
\begin{equation}\label{eqD:cond3}
\sqrt{4X+1} = w = \tfrac{1}{2}(1+4s)\left(1+\sqrt{1+\frac{16s}{(1+4s)^2}}\right)\,,
\end{equation}
where we used that $w\geq0$ to discard the negative root and which in turn leads to $X=g(p)^2$
\begin{equation}\label{eqD:cond4}
g(p)^2 = X = \tfrac{1}{16}(1+4s)^2\left(1+\sqrt{1+\frac{16s}{(1+4s)^2}}\right)^2 - \tfrac{1}{4}\,,
\end{equation}
so, finally, reinserting $s=\log(p)$:
\begin{equation}\label{eqD:g_pfinal}
g(p) = \tfrac{1}{2}\sqrt{\tfrac{1}{4}(1+4\log(p))^2\left(1+\sqrt{1+\frac{16\log(p)}{(1+4\log(p))^2}}\right)^2 - 1}\,,
\end{equation}
Inserting $k_\textrm{marg} = g(p)/\sqrt{\beta_2}$ back into the expression~\eqref{tauons_wave} for $\tau_\textrm{ons}(k)$ provides us with the marginal onset time 
\begin{equation}\label{eqD:tau_marg}
\tau_\textrm{marg} = \tau_\textrm{ons}(k_\textrm{marg})= \sqrt{\frac{g(p)^2}{\alpha_3(g(p)/\sqrt{\beta_2})^3}}\exp\left[-\frac{2\log(p)}{\sqrt{4g(p)^2 +1}-1}\right]\equiv h(p) \frac{\beta_2^{3/4}}{\alpha_3^{1/2}}\,,
\end{equation}
where we have defined the function
\begin{equation}\label{eqD:h_pfinal}
 h(p) = \frac{1}{g(p)^{1/2}}\exp\left[-\frac{2\log(p)}{\sqrt{4g(p)^2 +1}-1}\right]\,.
\end{equation}
By reinserting the expressions for $\alpha_3$ and $\beta_2$ from Eqs.~\eqref{eq:Adef}-\eqref{eq:Bdef} we obtain expressions for $\lambda_\textrm{marg}$ and $\tau_\textrm{marg}$ in physical quantities
\begin{eqnarray}
\lambda_\textrm{marg} &=& 2\pi \frac{m(\delta)}{g(p)}\,\frac{q_0}{U_0}\,,\label{eqD:lambdamarg}\\
\tau_\textrm{marg} &=& h(p)\,n(\delta)\,\left(\frac{\rho_l}{\sigma}\right)^{1/2}\left(\frac{q_0}{U_0}\right)^{3/2}\,,\label{eqD:taumarg}
\end{eqnarray}
where for notational convenience we have defined the following functions of the density ratio $\delta$
\begin{eqnarray}
m(\delta) &=& \sqrt{w(\delta)} = %\frac{\sqrt{\delta(1+\delta^2)}}{1-\delta^2} \approx \sqrt{\delta} + O(\delta^{5/2})\,,\label{eqD:m_delta}
\frac{\sqrt{\delta}}{1+\delta} \approx \sqrt{\delta} + O(\delta^{3/2})\,,\label{eqD:m_delta}\\
n(\delta) &=& (w(\delta))^{3/4}(1+\delta)^{1/2} = \frac{\delta^{3/4}}{1+\delta} %\frac{(\delta(1+\delta^2))^{3/4}}{(1-\delta)(1+\delta)^{1/2}} 
\approx \delta^{3/4} + O(\delta^{7/4})\,.\label{eqD:n_delta}
\end{eqnarray}\\

Finally, one may ask oneself if a similar procedure is also feasible to find the location of the maximum for the disk impact discussed in Subsection~\ref{subsec:diskimpact}. In that case, the condition $d\tau_\textrm{ons,d}/dk = 0$ leads to an equation of the form
\begin{equation}\label{eqD:diskcond1}
\beta_2k^2 = f\left(\log(p),\frac{\alpha_1\beta_2}{\alpha_3}\right)\,,
\end{equation}
with $\alpha_1$, $\alpha_3$ and $\beta_2$ as in Eqs.~\eqref{eq:Adef}-\eqref{eq:Bdef} and where determining the functional form of $f$ involves the solution of a fourth order polynomial equation. 

%%% APPENDIX E %%%

\section{Viscous solution}
\label{appE}

To determine the laminar solution $u(z,t)$ of a diverging gas flow which for large $z$ asymptotically tends to $U_g(\tau) = q_0/(U_0\tau)$, as sketched in Fig.~\ref{fig:appE}a, we may start from the well-known textbook solution of the impulsively started plate moving at speed $-u_0$ in the x-direction, where the Navier-Stokes equations are written down for a flow field of the form $\vec{V} = u(z,t)\,\vec{e}_x$, namely
\begin{equation}\label{eqE:NS}
\frac{\partial u}{\partial t} = \nu\frac{\partial^2 u}{\partial z^2} \,,
\end{equation}  
where $\nu$ is the kinematic gas viscosity and with boundary conditions $u(0,1) = 0$ and $u(\infty,t) = u_0$ in a reference frame that is moving with the plate (and in which the fluid above the plate appears to be moving with a velocity $+u_0$ in the positive $x$-direction). Initially, we have $u(z,0) = u_0$. This problem may be solved by introducing dimensionless variables $\zeta = z/\sqrt{\nu t}$ and $Y = u/u_0$, with which the above two-dimensional problem turns into a one-dimensional self-similar problem: $Y'' + \tfrac{1}{2}Y' = 0$, with $Y(0)=0$ and $Y(\infty)=1$, where $Y'$ and $Y''$ denote the first and second derivative of $Y$ with respect to $\zeta$. This equation is solved by an error function, $Y(\zeta) = \erf(\zeta/2)$, or, turning back to dimensional variables
\begin{equation}\label{eqE:solimpulsiveplate}
u(z,t) = u_0\erf\!\left[\frac{z}{2\sqrt{\nu t}}\right] \,.
\end{equation}  
Now in our problem with a diverging flow, we may write the gas velocity at any time $t_0$ smaller than the impact time $t_i$ as a superposition of small steps  
\begin{equation}\label{eqE:superpos}
U_g(t_0) = \int_{t=-\infty}^{t_0} \frac{dU_g}{dt}(t)dt = \sum_{k}  \frac{dU_g}{dt}(t_k)\Delta t_k = \sum_{k} \Delta U_g(t_k) \,,
\end{equation}  
where we adapted a rather sloppy notation, that disregards quite a number of convergence issues. Since the steps are increasing in size in time, and (of course) by virtue of the linearity of the governing equation \eqref{eqE:NS}, one may expect the resulting flow to be written as the superposition of the flow fields generated by each of the small steps, i.e.
\begin{equation}\label{eqE:superposfield1}
u(z,t_0) =  \sum_{k} \Delta U_g(t_k)\erf\!\left[\frac{z}{2\sqrt{\nu (t_0-t_k)}}\right] =   \int_{t=-\infty}^{t_0} \frac{dU_g}{dt}(t)\erf\!\left[\frac{z}{2\sqrt{\nu (t_0-t)}}\right] dt  \,.
\end{equation} 
 Now we write $U_g(t) = q_0/(U_0(t_i-t))$, from which $dU_g/dt = q_0/(U_0(t_i-t)^2)$, such that
 \begin{equation}\label{eqE:superposfield2}
u(z,t_0) =  \frac{q_0}{U_0} \int_{t=-\infty}^{t_0} \frac{1}{(t_i-t)^2}\erf\!\left[\frac{z}{2\sqrt{\nu (t_0-t)}}\right] dt  \,,
\end{equation} 
or, dividing by $U_g(t_0) = q_0/(U_0(t_i-t_0))$
 \begin{equation}\label{eqE:superposfield3}
\frac{u(z,t_0)}{U_g(t_0)}  =  \int_{t=-\infty}^{t_0} \frac{(t_i-t_0)}{(t_i-t)^2}\erf\!\left[\frac{z}{2\sqrt{\nu (t_0-t)}}\right] dt  \,,
\end{equation} 
which with $\tau_0 = t_i-t_0$ and $s \equiv t_0-t$ can be written as
\begin{equation}\label{eqE:superposfield4}
\frac{u(z,t_0)}{U_g(t_0)} = \int_{s=0}^{\infty} \frac{\tau_0}{(\tau_0 + s)^2}\,\,\textrm{erf}\!\left[\frac{z}{2\sqrt{\nu s}}\right]ds\,.
\end{equation}
Now, defining $\eta_0 \equiv z/\sqrt{\nu \tau_0}$ and $w \equiv z/\sqrt{\nu s}\,\,\Rightarrow\,\,s = z^2/(\nu w^2)$, with which, $ds = -2z^2/(\nu y^3)$ we obtain
\begin{eqnarray}
\frac{u(z,t_0)}{U_g(t_0)} &=& \frac{2z^2}{\nu\tau_0}\int_{w=0}^{\infty} \frac{1}{(1 + (z^2/\nu\tau_0)/w^2)^2w^3}\,\,\textrm{erf}\!\left[\tfrac{1}{2}w\right]dw\nonumber\\
&=& \eta_0^2\int_{w=0}^{\infty} \frac{1}{(1 + (\eta_0/w)^2)^2w^3}\,\,\textrm{erf}\!\left[\tfrac{1}{2}w\right]dw \,,\label{eqE:superposfield5}
\end{eqnarray}
Clearly, with this last expression we observe that the ratio $u(z,t_0)/U_g(t_0)$ may be written as a function of one combined variable only, namely $\eta_0 = z/\sqrt{\nu\tau_0}$, with which 
\begin{equation}\label{eqE:funcform}
\frac{u(z,t)}{U_g(t)} = F\left(\frac{z}{\sqrt{\nu\tau}}\right)\,,
\end{equation}
where from hereon we have dropped the now superfluous subscript zero in $\tau_0$ and $\eta_0$. Most specifically, we may collapse profiles $u(z,t)$ for different $t$ onto a single master curve given by Eq.~\eqref{eqE:funcform}, and as a consequence the typical boundary layer profile width $\delta_\textrm{BL}(\tau)$ is proportional to $\sqrt{\nu\tau}$:
\begin{equation}\label{eqE:BLwidthsim}
\delta_\textrm{BL}(\tau) \sim \sqrt{\nu\tau}\,.
\end{equation}  
It is remarkable to see that for this particular problem the viscous boundary layer thickness in fact \emph{decreases} with time. Intuitively, this behavior may be understood from the fact that the gas flow speed $U_g(\tau)$ is rapidly increasing when $\tau\to 0$, which implies that for equal time steps $\Delta t$ the velocity steps $\Delta U_g = (dU_g/dt)\Delta t$ become larger and larger, whereas the developing viscous boundary layer is trying to keep up with these increasing steps.

A last transformation $\xi = \eta/w\,\,\Rightarrow \,\, w = \eta/\xi\,\,\Rightarrow \,\,dw = -\eta/\xi^2 d\xi$, brings the expression \eqref{eqE:superposfield5} in an integrable form
\begin{equation}\label{eqE:superposfield6}
\frac{u(z,t)}{U_g(t)} = \int_{\xi=0}^{\infty} \frac{\xi}{(1 + \xi^2)^2}\,\,\erf\!\left[\frac{\eta}{2\xi}\right]d\xi \,.
\end{equation}
This integral can be evaluated analytically as
\begin{equation}\label{eqE:result}
\frac{u(z,t)}{U_g(t)} = F(\eta) = \tfrac{1}{2}\sqrt{\pi}\,\eta\,\exp\!\left[\frac{\eta^2}{4}\right]\erfc\!\left[\frac{\eta}{2}\right] \,,
\end{equation}
where $\erfc(x) = 1 - \erf(x)$ is the complementary error function. This result is plotted in Fig.~\ref{fig:appE}b. Now that we have obtained an analytic expression for the flow field we may define the boundary layer thickness $\delta_\textrm{BL}(\tau)$ as the length scale for which the velocity attains 99\% of the asymptotic value, i.e., $u(z,t)/U_g(t) = F(\eta_\textrm{BL}) = 0.99$ leading to $\eta_\textrm{BL} = 13.93$, from which
\begin{equation}\label{eqE:BLwidth}
\delta_\textrm{BL}(\tau) = 13.93 \sqrt{\nu\tau}\,.
\end{equation}  
This value $\eta_\textrm{BL} = 13.93$ is indicated by the horizontal dashed red line in Fig.~\ref{fig:appE}b. 
 
Finally, it is good to realize that the expression~\eqref{eqE:result} is a solution of the (dimensionless) boundary value problem for $F(\eta)$
\begin{equation}
F'' -\tfrac{1}{2}\eta F' - F = -1\,, \qquad F(0) = 0\,,\quad F(\infty) = 1\,,\label{eqE:BVP}
\end{equation}  
where (again) $F'$ and $F''$ denote the first and second derivative of $F$ with respect to $\eta$. This boundary value problem could be obtained by dimensional analysis. Arguing that (as soon as memory from any start-up effects have been erased) the problem is fully described by the dimensional quantities $u(z,\tau)$, $U_g(\tau)$, $\nu$, $z$, and $\tau$, this leads to a description in terms of two dimensionless variables $Y = u(z,\tau)/U_g(\tau)$ and $\eta = z/\sqrt{\nu\tau}$. Now, naively starting from the equation of motion~\eqref{eqE:NS} with boundary conditions $u(0,\tau)=0$ and $u(\infty,\tau)=U_g(\tau)$ would have lead to the slightly different boundary value problem $Y'' -\tfrac{1}{2}\eta Y' - Y = 0$,  $Y(0) = 0$, $Y(\infty) = 1$ which does not possess a bounded solution. 

One does obtain the correct boundary value problem if one realizes that from the perspective of someone moving with the accelerating gas flow, the laboratory frame of reference is a non-inertial frame of reference where there is an inertial acceleration corresponding to $a_{i} = dU_g/dt = U_g/\tau$, that is, Eq.~\eqref{eqE:NS} needs to be replaced by
\begin{equation}
\frac{\partial u}{\partial t} = \nu\frac{\partial^2 u}{\partial z^2} + \frac{U_g}{\tau}\,.\label{eqE:NSrev}
\end{equation}  
Starting from the above equation in the derivation leads to the boundary value problem~\eqref{eqE:BVP}.

\bibliographystyle{jfm}
\bibliography{TimedependentKH}

\end{document}